\pdfoutput=1
\documentclass[pra,nofootinbib,twocolumn,amsmath,amssymb,superscriptaddress]{revtex4-2}
\usepackage{graphicx,amsmath,relsize,epstopdf,color,mathtools,bm, braket,rotating,amssymb, physics}
\usepackage[hyphenbreaks]{breakurl}
\usepackage[colorlinks=true,linkcolor=blue,citecolor=blue,urlcolor =blue]{hyperref}
\usepackage[normalem]{ulem}
\usepackage{booktabs}
\usepackage[table,xcdraw]{xcolor}
\usepackage{gensymb}

\newcommand{\RE}{\mathop{\mathrm{Re}} \nolimits}

\newcommand{\betac}{|\mathring{\beta}|}
\newcommand{\notJ}{{J\mkern-8mu{\backslash}}}
\usepackage{soul,xcolor}
\usepackage{mathtools}
\usepackage{graphicx}
\usepackage{dsfont}
\usepackage{placeins}
\newcommand{\vac}{\text{vac}}
\newcommand{\sinc}{\text{sinc}}
\renewcommand{\ol}[1]{\overline{#1}}
\usepackage{soul}
\usepackage{cancel}
\usepackage[utf8]{inputenc}
\DeclareUnicodeCharacter{0248}{\nJ}
\usepackage{tikz}
\usepackage{amsmath}

\newcommand{\stkout}[1]{\ifmmode\text{\sout{\ensuremath{#1}}}\else\sout{#1}\fi}

\begin{document}
	\setstcolor{red}	
	
	\title{Photodetection of Squeezed Light: a Whittaker-Shannon Analysis}
	
	\author{Jasper Kranias} 
    \affiliation{Department of Physics, University of Toronto, 60 St. George Street, Toronto, ON M5S 1A7}
    \author{Christian Drago}
    \affiliation{Department of Physics, University of Toronto, 60 St. George Street, Toronto, ON M5S 1A7}  
    \affiliation{Xanadu, Toronto, ON M5G 2C8, Canada}
    \author{Colin Vendromin}
    \affiliation{Department of Physics, University of Toronto, 60 St. George Street, Toronto, ON M5S 1A7}  
    \affiliation{Xanadu, Toronto, ON M5G 2C8, Canada}
    \author{J. E. Sipe}
    \affiliation{Department of Physics, University of Toronto, 60 St. George Street, Toronto, ON M5S 1A7}



	\begin{abstract}
    The Whittaker-Shannon decomposition provides a temporally localized description of squeezed light, making it applicable in the CW limit and leading to a definition of squeezing strength based on the number of photon pairs at a time. We show examples of its usefulness by calculating quadrature variance in a homodyne detection scheme, coincidence detection probabilities in the continuous-wave limit, and analyzing the Hong-Ou-Mandel effect for strongly squeezed light. Quadrature uncertainty falls farther below the shot noise limit when squeezing is strong, but effects due to correlations between photon pairs are most significant with weak squeezing. Our analysis extends previous results to more general scenarios, and we leverage the Whittaker-Shannon formalism to interpret them based on the temporal properties of photon pairs.
    \end{abstract}
	
	\maketitle

\section{Introduction}

Squeezed states of light can exhibit interesting spectro-temporal behaviour that does not arise for ordinary light \cite{Glauber}, giving rise to applications in quantum computing \cite{Bourassa2021blueprintscalable,Xanadu_GKP}, quantum sensing \cite{XanaduSense}, high-precision optical interferometry \cite{Yurke}, and quantum key distribution \cite{TAKESUE2010276}. Some of these require strongly squeezed light, and in those that utilize single photon pairs it is important to account for the chance of generating multiple pairs \cite{TAKESUE2010276,Takeoka_2015,Adam_Multiphoton_Noise,Kim_QKD,Brewster_polarization_entanglement_degradation}. Therefore, a full analysis of squeezed light requires going beyond the single pair regime \cite{Quesada:22}, often done by taking a Schmidt decomposition of the joint amplitude. However, the Schmidt decomposition is less applicable to continuous-wave (CW) squeezed light, since the Schmidt number diverges. In the Whittaker-Shannon formalism, squeezed light is described instead using a set of modes localized in time \cite{drago2023takingapartsqueezedlight}, allowing it to be applied in the CW limit since only the modes near the times of interest need be considered. The Whittaker Shannon formalism also leads to a definition of squeezing strength based on how many photon pairs exist near a given time; a very long pulse of squeezed light containing many photons can be thought of as weakly squeezed in the sense that there is only one photon pair at a time. 


This work utilizes the Whittaker-Shannon decomposition to investigate the behaviour of squeezed light in multiple detection schemes, generalizing some results that previously were only known in the case of finite pulses to the CW limit, and providing an analysis of how squeezing strength affects the detection statistics of squeezed light. Throughout the paper, we use the Whittaker-Shannon formalism to build intuition on the behaviour of squeezed light based on the temporal correlations of photon pairs and the rate they are generated.

In Section \ref{Formalism Section} we outline the Whittaker-Shannon decomposition for nondegenerate squeezed light, show how it leads to a description of the state within a time window, and make comparisons with the usual method of discretizing the temporal modes. In the process of building the Whittaker-Shannon formalism, we produce a disentangling formula for the multimode nondegenerate squeezing operator. To our knowledge, this has not been correctly derived before; Ref. \cite{Puri} provides a formula disagreeing with ours, though it appears not to be used elsewhere in the literature. In Section \ref{Homodyne Sec} we formulate homodyne detection in terms of the Whittaker-Shannon decomposition both in the continuous-wave (CW) limit and for finite pulses. In both cases, the quadrature variance is minimized further with increased squeezing strength, showing that our parameter for squeezing strength corresponds well with traditional measures \cite{Strong_Optomechanical_Squeezing,Intense_Squeezing,10dB_strength}. In Section \ref{Coincidence Sec} we extend Takesue's work \cite{TAKESUE2010276} on the coincidence visibilities of a polarization dependent detection scheme to a more general description of the squeezed ket. We find expressions for the photon number and coincidence probabilities in a time window by writing projection operators for detection inside the time window in terms of the Whittaker-Shannon modes within.
In Section \ref{HOM Sec} we derive expressions for the coincidence probability in a Hong-Ou-Mandel scheme for multimode squeezed. Unlike previous expressions that either only account for some higher order terms \cite{Ferrari_HOM}, or are not applicable in the CW limit \cite{Takeoka_2015}, our formulae can be applied to CW squeezed light of arbitrary squeezing strength. The presence of multiple photon pairs reduces the visibility of the Hong-Ou-Mandel dip, agreeing with previous work \cite{Takeoka_2015,Zhang_2010_HOM_experiment} In contrast to homodyne detection where quantum effects were most significant for strong squeezing, the schemes of Sections \ref{Coincidence Sec} and \ref{HOM Sec} display greater effects with weak squeezing.





\section{Whittaker-Shannon Formalism}
\label{Formalism Section}
The Whittaker-Shannon decomposition describes squeezed light in terms of novel supermodes that are localized in time \cite{drago2023takingapartsqueezedlight}. In this section, we outline the decomposition for nondegenerate squeezed light, and show how we can approximate the state in certain time windows using only the nearby supermodes. 
From consideration of a finite time window, we provide a definition of weak squeezing that reflects the density of photon pairs within a range of time; ``weakly squeezed'' can then describe a ket in the CW limit, even though there are an infinite number of pairs. We also find the $\boldsymbol{N}$ and $\boldsymbol{M}$ moments in terms of the Whittaker-Shannon formalism, and compare the Whittaker-Shannon decomposition to a typical method of discretizing the temporal modes.

\subsection{Whittaker-Shannon Decomposition}

The Whittaker-Shannon decomposition was first introduced for degenerate squeezed light \cite{drago2023takingapartsqueezedlight}. The notation for the degenerate regime can be used generally, but it will be convenient for us to introduce an explicit notation for the nondegenerate regime since commutativity of some of the operators simplifies certain calculations. Here we consider light propagation in one-direction, as in an optical fiber or waveguide mode. If the frequency ranges of $\omega_1$ and $\omega_2$ over which the joint spectral amplitude $\gamma(\omega_1,\omega_2)$ is significant are far apart, we identify the first range with the ``signal'', and the second with the ``idler'', and then assign separate operators (with different center frequencies) to the signal and idler ranges. After shifting $\gamma(\omega_1,\omega_2)$ so that the center frequencies of the signal and idler modes are at zero, we can write a nondegenerate squeezed ket as
\begin{equation}
\label{nondegket}
    \ket{\psi}=e^{\beta\int d\omega_1d\omega_2\gamma(\omega_1,\omega_2)a^\dagger(\omega_1)b^\dagger(\omega_2)-h.c.}\ket{\vac},
\end{equation}
with
\begin{equation}
\begin{aligned}
\label{Commutators}
    [a(\omega),b^\dagger(\omega')]&=0,\\ [a(\omega),a^\dagger(\omega')]&=\delta(\omega-\omega'),\\ [b(\omega),b^\dagger(\omega')]&=\delta(\omega-\omega'),
\end{aligned}
\end{equation}
and $\ket{\vac}$ the vacuum state \cite{drago2023takingapartsqueezedlight,Quesada:22}. We adopt the convention that integrals range from $-\infty$ to $\infty$ unless indicated otherwise. Eq.~\eqref{nondegket} can also apply if the signal and idler photons share the same center frequency, but are either spatially separated, or are associated with different transverse modes. The joint amplitude is not symmetric in general ($\gamma(\omega_1,\omega_2)\neq\gamma(\omega_2,\omega_1)$), and is normalized by
\begin{equation}
    \int d\omega_1 d\omega_2|\gamma(\omega_1,\omega_2)|^2=1.
\end{equation}
Equivalently, we can write the squeezed ket in terms of the joint temporal amplitude
\begin{equation}
    \overline{\gamma}(t_1,t_2)=\int\frac{d\omega_1d\omega_2}{2\pi}e^{-i\omega_1t_1}e^{-i\omega_2t_2}\gamma(\omega_1,\omega_2)
\end{equation}
and the Fourier transforms of the annihilation operators
\begin{equation}
    \begin{aligned}
        \ol{a}(t)&=\int \frac{d\omega}{\sqrt{2\pi}} e^{-i\omega t}a(\omega), & \ol{b}(t)&=\int \frac{d\omega}{\sqrt{2\pi}} e^{-i\omega t}b(\omega),
    \end{aligned}
\end{equation}
as
\begin{equation}
\label{nondegket_temporal}
    \ket{\psi}=e^{\beta\int dt_1dt_2\ol{\gamma}(t_1,t_2)\ol{a}^\dagger(t_1)\ol{b}^\dagger(t_2)-h.c.}\ket{\vac}.
\end{equation}
We refer to the modes described by $\ol{a}(t)$ and $\ol{b}(t)$ as continuous temporal (CT) modes, and they obey commutation relations like Eq.~\eqref{Commutators}. We take the ket in Eq.~\eqref{nondegket_temporal} to identify the state at $t=0$. The state at time $t$ is then
\begin{equation}
    \ket{\psi(t)}=e^{\beta\int d\omega_1d\omega_2\gamma(\omega_1,\omega_2)\exp(i\omega_1t+i\omega_2t)a^\dagger(\omega_1)b^\dagger(\omega_2)-h.c.}\ket{\vac},
\end{equation}
or in terms of the temporal joint amplitude,
\begin{equation}
    \ket{\psi(t)}=e^{\beta\int dt_1dt_2\ol{\gamma}(t_1,t_2)\ol{a}^\dagger(t_1-t)\ol{b}^\dagger(t_2-t)-h.c.}\ket{\vac}.
\end{equation}
Since the light propagates with velocity $v$ --- we neglect group velocity dispersion --- $\ol{a}(t_1-t)$ represents the field at $t_1$ at a distance $d=vt$ from the pulse center at time zero \cite{drago2023takingapartsqueezedlight}. Therefore, the operators in Eqs.~\eqref{nondegket} and \eqref{nondegket_temporal} represent the field at the detectors if they lie at position $d$ and we shift the origin of time by $t=d/v$.

If $\gamma(\omega_1,\omega_2)$ is approximately bandwidth limited by $\Omega$, significant only when 
\begin{equation}
\label{bandlimit}
    -\frac{\Omega}{2}\leq\omega_1,\omega_2\leq\frac{\Omega}{2},
\end{equation}
then we can perform a Whittaker-Shannon decomposition of the joint amplitude \cite{drago2023takingapartsqueezedlight} to write
\begin{equation}
\label{Whittaker-Shannon Ket}
    \ket{\psi}=S\ket{\text{vac}},
\end{equation}
where 
\begin{equation}
    S=e^{\sum_{n,m}\beta_{nm}A_n^\dagger B_m^\dagger-h.c.}
\end{equation}
is the nondegenerate squeezing operator, $\beta_{nm}=\beta\tau\overline{\gamma}(n\tau,m\tau)$, $\tau=2\pi/\Omega$
and we define
\begin{equation}\begin{aligned}
\label{WS_Operators}
    A^\dagger_n&=\int dt\overline{\chi}_n(t)\overline{a}^\dagger(t),\\ B^\dagger_n&=\int dt\overline{\chi}_n(t)\overline{b}^\dagger(t),
\end{aligned}\end{equation}
where
\begin{equation}
    \overline{\chi}_n(t)=\frac{1}{\sqrt{\tau}}\sinc\bigg(\frac{\pi (t-n\tau)}{\tau}\bigg).
\end{equation}
The Whittaker-Shannon timescale $\tau$ is typically on the order of the coherence time, the range of $|t_2-t_1|$ over which $|\ol{\gamma}(t_1,t_2)|^2$ is significant. 
Unlike for degenerate squeezing, $\beta_{nm}$ is not in general a symmetric matrix \cite{drago2023takingapartsqueezedlight}. The Whittaker-Shannon modes $\{\overline{\chi}_n(t)\}$ are orthonormal \cite{drago2023takingapartsqueezedlight}, which guarantees that the supermode operators obey the usual commutation relations
\begin{equation}
\label{WS_operator_commutators}
    [A_n,B^\dagger_m]=0,\quad [A_n,A^\dagger_m]=\delta_{nm},\quad [B_n,B^\dagger_m]=\delta_{nm}.
\end{equation}
The supermodes associated with the operators $A^\dagger_n$ and $B^\dagger_n$ are localized around $n\tau$, in the sense that $A^\dagger_n$ and $B^\dagger_n$ are composed mostly of $\overline{a}^\dagger(t)$ or $\overline{b}^\dagger(t)$ near $t=n\tau$. 
Since $\{\overline{\chi}_n(t)\}$ forms a complete set for expanding functions whose Fourier transforms are nonzero only for frequencies satisfying Eq.~\eqref{bandlimit} \cite{drago2023takingapartsqueezedlight}, and we are assuming that only these frequencies are necessary to describe the state, we can invert Eq.~\eqref{WS_Operators} to write the CT modes as
\begin{equation}\begin{aligned}
\label{invertedWSmodes}
    \overline{a}(t)&=\sum_n\overline{\chi}_n(t)A_n, & \overline{b}(t)&=\sum_n\overline{\chi}_n(t)B_n.
\end{aligned}\end{equation}
If we let $\ol{\gamma}_{max}=max(|\ol{\gamma}(n\tau,m\tau)|)$ and define
\begin{equation}
    \mathring{\beta}=\beta\tau\ol{\gamma}_{max},
\end{equation}
we can then write $\beta_{nm}$ as
\begin{equation}
\label{betac_def}
    \beta_{nm}=\mathring{\beta}r_{nm},
\end{equation}
where $r_{nm}=\ol{\gamma}(n\tau,m\tau)/\ol{\gamma}_{max}$, with $|r_{nm}|\leq1$. Since $\betac$ is the maximum of $|\beta_{nm}|$, it sets the magnitude of the matrix $\boldsymbol{\beta}$ and other matrices we will define later. When $\betac$ is small we can expand functions of those matrices to low orders. We argue in Section~\ref{WeakKetSec} that $\betac$ quantifies the squeezing strength.

We shall now distinguish the Whittaker-Shannon decomposition from another way of decomposing the ket in terms of temporally localized modes. Consider discretizing the CT mode operators into time bins of size $T_D$ by defining
\begin{equation}
\label{DT mode operators}
\begin{aligned}
    \ol{a}_n&=\frac{1}{\sqrt{T_D}}\int_{D_n}dt\ol{a}(t), & \ol{b}_m&=\frac{1}{\sqrt{T_D}}\int_{D_m}dt\ol{b}(t), 
\end{aligned}
\end{equation}
where $D_n$ indicates that the integral ranges from $(n-\frac{1}{2})T_D$ to $(n+\frac{1}{2})T_D$. We refer to this as a \textit{standard discretization} into discrete temporal (DT) modes, and the DT mode operators $\ol{a}_n$ and $\ol{b}_m$ obey commutation relations like Eq.~\eqref{WS_operator_commutators}. If we partition the integrals into time bins of size $T_D$ and approximate the joint temporal amplitude in each time bin as taking the value at the center by assuming $\ol{\gamma}(t_1,t_2)$ varies slowly over $T_D$, we have
\begin{equation}
    \begin{aligned}
        &\beta\int dt_1dt_2\ol{\gamma}(t_1,t_2)\ol{a}^\dagger(t_1)\ol{b}^\dagger(t_2)\\
        &=\sum_{n,m}\beta T_D\ol{\gamma}(nT_D,mT_D)\ol{a}^\dagger_n\ol{b}^\dagger_m.
    \end{aligned}
\end{equation}
The squeezed ket then takes the same form as a Whittaker-Shannon decomposition with $\tau=T_D$:
\begin{equation}
\label{standard_discretization_decomp}
    \ket{\psi}=e^{\sum_{n,m}\beta_{nm}\ol{a}^\dagger_n\ol{b}^\dagger_m-h.c.},
\end{equation}
but the Whittaker-Shannon decomposition remains distinct from this method, and holds some advantages over it. First, the Whittaker-Shannon decomposition allows for a timescale as large as the coherence time, where $\ol{\gamma}(t_1,t_2)$ is not slowly varying, so it requires less terms than Eq.~\eqref{standard_discretization_decomp} to cover the same time window. A timescale on the order of the coherence time also leads to a natural definition of squeezing strength. Moreover, Eq.~\eqref{DT mode operators} cannot be inverted to write the CT mode operators in terms of the DT mode operators, so we cannot use the standard discretization to calculate quantities with explicit time dependence (such as the moments in Eq.~\eqref{nondegen moments}). However, many calculations, such as those in Sections \ref{Coincidence Sec} and \ref{HOM Sec}, are defined purely in terms of projection onto the Whittaker-Shannon supermode operators. Since Eq.~\eqref{standard_discretization_decomp} takes the same form as the Whittaker-Shannon decomposition, a calculation defined by projection of Eq.~\eqref{standard_discretization_decomp} onto DT mode operators will have an identical result to the corresponding calculation based on the Whittaker-Shannon decomposition.

\subsection{Partitioning the Ket in Time}
\label{WeakKetSec}

A model often used to qualitatively represent squeezed light is the double Gaussian joint amplitude:
\begin{equation}
    \begin{aligned}
        \gamma(\omega_1,\omega_2)&=\sqrt{\frac{T_pT_c}{\pi^2}}e^{-\frac{T_c^2(\omega_1-\omega_2)^2}{4\pi}}e^{-\frac{T_p^2(\omega_1+\omega_2)^2}{4\pi}},\\
        \ol{\gamma}(t_1,t_2)&=\sqrt{\frac{1}{T_pT_c}}e^{-\frac{\pi(t_1-t_2)^2}{4T_c^2}}e^{-\frac{\pi(t_1+t_2)^2}{4T_p^2}}.
    \end{aligned}
\end{equation}
The length of the pulse is characterized by $T_p$, and $T_c<T_p$ can be identified as a coherence time. The inverse of the coherence time $B_c=1/T_c$ identifies the bandwidth of the joint spectral amplitude \cite{Quesada:22,DragoHOM}.

\begin{figure}
    \centering
    \includegraphics[width=0.45\textwidth]{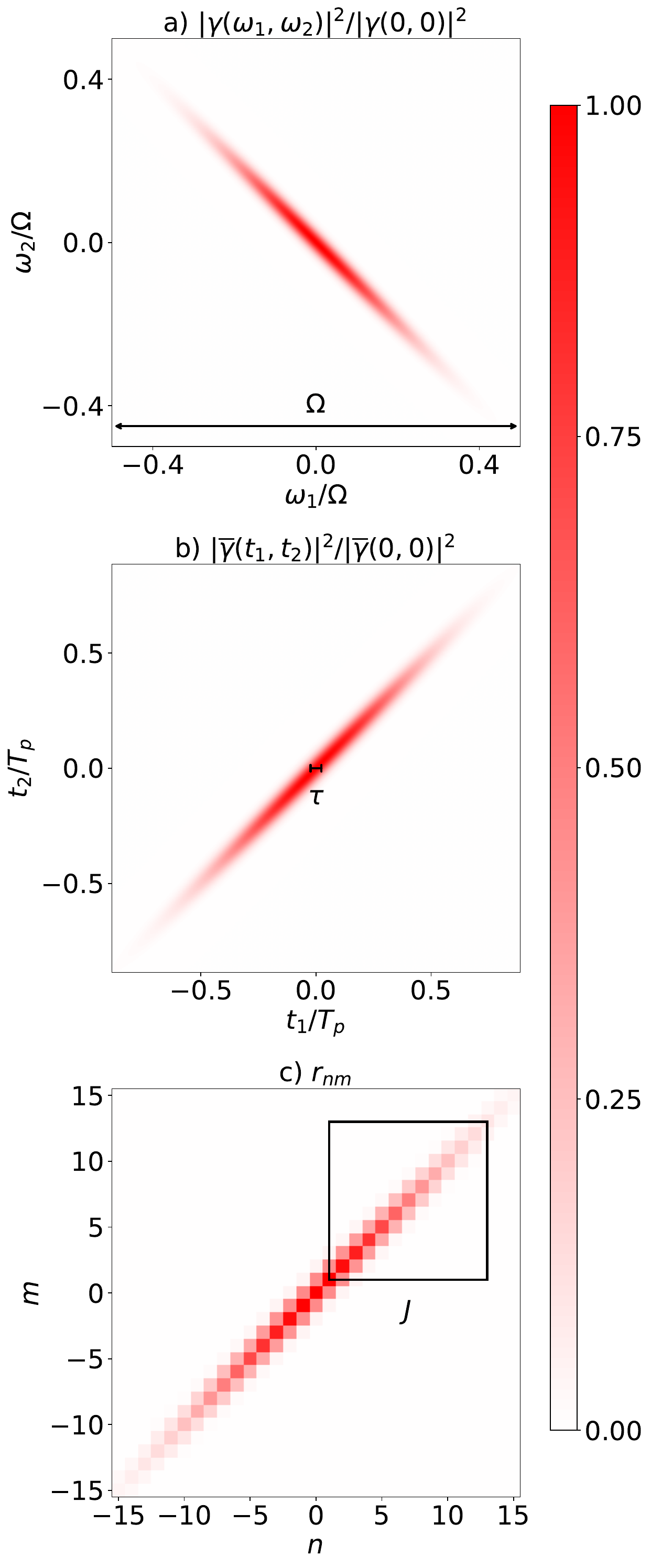}
    \caption{a) Normalized double-Gaussian joint spectral amplitude $|\gamma(\omega_1,\omega_2)|^2$ with axes normalized by $\Omega$. b) Normalized double-Gaussian joint temporal intensity $|\ol{\gamma}(t_1,t_2)|^2$ with axes normalized by $T_p$. c) Amplitudes $r_{nm}$ of the Whittaker-Shannon decomposition of the double-Gaussian joint amplitude. Observe that $r_{nm}$ is small when more than one space away from the diagonal. The black square represents the nonzero elements of $\boldsymbol{\beta}^J$ we could take if we were interested in times close to $t_J=7\tau$; any $r_{nm}$ outside of the box is either small, or has both $n\tau$ and $m\tau$ far from $t_J$. These plots were made with $T_p/T_c=15$.}
    \label{DG Fig}
\end{figure}

It is natural to choose a bandlimit on the order of the bandwidth; we take $\Omega=2\pi/T_c$ and $\tau=T_c$. In Fig.~\ref{DG Fig} we show the intensity of the spectral and temporal double Gaussian joint amplitudes and the matrix $r_{nm}$ of its Whittaker-Shannon decomposition.
Our choice of $\Omega$ leads to
\begin{equation}
    r_{nm}=e^{-\frac{\pi(n-m)^2}{4}}e^{-(\frac{T_c}{T_p})^2\frac{\pi(n+m)^2}{4}},
\end{equation}
and we observe that $r_{nm}$ is small far from the diagonal; off-diagonal terms will be a factor of at least $e^{-\pi}\approx0.04$ smaller than the maximum for $|n-m|\geq2$. In the CW limit, $T_p\to\infty$ \cite{Absorbtion_of_squeezed_light_CW_lim}, so we take
\begin{equation}
    r_{nm}\to e^{-\frac{\pi(n-m)^2}{4}},
\end{equation}
and can still use Eq.~\eqref{betac_def} since $\betac$ remains finite in the CW limit even though $\beta$ diverges \cite{drago2023takingapartsqueezedlight}. Thus, the Whittaker-Shannon decomposition allows us to go to the CW limit analytically.

For the double Gaussian and other joint amplitudes with the property that $\ol{\gamma}(t_1,t_2)$ is small when $t_2$ is far from $t_1$, we can follow earlier arguments \cite{drago2023takingapartsqueezedlight} to approximate the ket locally near a particular time. If we are only interested in the squeezed light near some time $t_J$, then we can partition the matrix $\boldsymbol{\beta}$ into
\begin{equation}
    \boldsymbol{\beta}=\boldsymbol{\beta}^J+\boldsymbol{K}.
\end{equation}
The matrix $\boldsymbol{\beta}^J$ is centered at an index $n_J=[\frac{t_J}{\tau}]$, where $[\cdot]$ denotes the nearest integer, and we choose $d_J$ such that $\beta_{n_J,m}$ ($\beta_{n,n_J}$) can be neglected for $|n_J-m|>\frac{d_J}{2}$ ($|n-n_J|>\frac{d_J}{2}$). We take $\beta^J_{nm}$ to be equal to $\beta_{nm}$ when $n,m$ are within $(d_J-1)/2$ of $n_J$ and zero everywhere else; Fig.~\ref{DG Fig}c) shows a schematic of such a partition. The partition size $d_J$ should be larger than the range of times in which we are interested, since the statistics at times close to the edge of the partition will be affected by elements of $\boldsymbol{\beta}$ not included in $\boldsymbol{\beta}^J$. 
When the coherence time is sufficiently smaller than the pulse duration, $d_J$ can be chosen so that the time window is smaller than the pulse, and we can describe the state near $t_J$ using only parts of $\boldsymbol{\beta}$ that correspond to times near $t_J$. However, this is not possible for all joint amplitudes. For example, if we rotated the double Gaussian in Fig.~\ref{DG Fig} b) by $45\degree$ in the $t_1-t_2$ plane, then $|\ol{\gamma}(t_1,t_2)|^2$ would be significant over the entire range of significant $t_2$ for most significant values of $t_1$ and we could not choose a $d_J$ that allows us to describe the state near $t_J$ without considering the whole pulse. Later on, we derive a few results in terms of $\boldsymbol{\beta}^J$, and if the time window $J$ is shorter than the pulse, then these expressions are only valid for joint amplitudes that are small far from the diagonal. However, the results apply to any joint amplitude if $J$ covers the entire pulse so that $\boldsymbol{\beta}^J=\boldsymbol{\beta}$.

For the times of interest we can now approximate the squeezing operator as
\begin{equation}
    S\approx e^{\sum^J_{n,m}\beta^J_{nm}A^\dagger_nB^\dagger_m-h.c.}\equiv S_J.
\end{equation}
The superscript $J$ on the sum indicates that we only sum over the nonzero elements of $\boldsymbol{\beta}^J$, and if we let $\ket{\vac}^J$ be the vacuum state corresponding to the Whittaker-Shannon modes with indices inside the time window, the state near $t_J$ can be approximately represented by
\begin{equation}
\label{LocalKet_not_Disentangled}
    \ket{\psi_J}=S_J\ket{\vac}^J=e^{\sum^J_{n,m}\beta^J_{nm}A^\dagger_nB^\dagger_m-h.c.}\ket{\vac}^J.
\end{equation}

Consider the left and right polar decompositions of $\boldsymbol{\beta}^J$,
\begin{equation}\begin{aligned}
\label{PolarDecomp}
    \boldsymbol{\beta}^J&=\boldsymbol{U}^J\boldsymbol{P}^J, & \boldsymbol{\beta}^J&=\boldsymbol{Q}^J\boldsymbol{U}^J,
\end{aligned}\end{equation}
respectively, where $\boldsymbol{P}^J=\sqrt{(\boldsymbol{\beta}^J)^\dagger\boldsymbol{\beta}^J}$, $\boldsymbol{Q}^J=\sqrt{\boldsymbol{\beta}^J(\boldsymbol{\beta}^J)^\dagger}=\boldsymbol{U}^J\boldsymbol{P}^J(\boldsymbol{U}^J)^\dagger$, and $\boldsymbol{U}^J$ is unitary. To gain some intuition on the state we use the disentangled form of the nondegenerate squeezing operator (Appendix \ref{DisentanglingDix})
\begin{equation}
\begin{aligned}
    S_J&=|\boldsymbol{W}^J|e^{\sum^J_{n,m}T^J_{nm}A^\dagger_nB^\dagger_m}e^{\sum^J_{n,m}(L^J_{nm}A^\dagger_nA_m+Y^J_{nm}B^\dagger_nB_m)}\\
    &\quad\times e^{-\sum^J_{n,m}V^J_{nm}A_nB_m},
\end{aligned}
\end{equation}
where
\begin{equation} \begin{aligned}
\label{Disentangling Matrices}
    \boldsymbol{W}^J&=\sech\boldsymbol{Q}^J\\
    \boldsymbol{T}^J&=(\tanh\boldsymbol{Q}^J)\boldsymbol{U}^J\\
    \boldsymbol{L}^J&=\ln(\sech\boldsymbol{Q}^J)\\
    \boldsymbol{Y}^J&=\ln(\sech(\boldsymbol{P}^J)^T)\\
    \boldsymbol{V}^J&=((\boldsymbol{U}^J)^\dagger(\tanh\boldsymbol{Q}^J))^T, 
\end{aligned} \end{equation}
to find
\begin{equation}
\label{LocalKet}
    \ket{\psi_J}=|\boldsymbol{W}^J|e^{\sum^J_{n,m}T^J_{nm}A^\dagger_nB^\dagger_m}\ket{\vac}^J.
\end{equation}
If we are in the limit of weak squeezing ($|\mathring{\beta}|\ll1$), the mean number of pairs in $\ket{\psi_J}$ is
\begin{equation}
    N_J\equiv\int dt\bra{\psi_J}\ol{a}^\dagger(t)\ol{a}(t)\ket{\psi_J}\approx\Tr((\boldsymbol{Q}^J)^2)+\mathcal{O}(|\mathring{\beta}|^4).
\end{equation}
To leading order, $N_J$ is proportional to $\betac^2$ since $\boldsymbol{Q}^J$ is proportional to $\betac$. However, $|\mathring{\beta}|\ll1$ does not guarantee that $N_J\ll1$ since the trace can be large if $\dim(\boldsymbol{Q}^J)$ is large. If $N_J\ll1$, we can expand Eq.~\eqref{LocalKet} to first order in $|\mathring{\beta}|$ to find
\begin{equation}
\label{WeakKet}
    \ket{\psi_J}\approx|\boldsymbol{W}^J|(\ket{\vac}^J+\sqrt{N_J}\ket{II}_J),
\end{equation}
where $\ket{II}$ is the normalized two-photon ket
\begin{equation}
\label{TwoPhotonState}
    \ket{II}_J=\frac{1}{\sqrt{N_J}}\sum^J_{n,m}T^J_{nm}A^\dagger_nB^\dagger_m\ket{\vac}^J,
\end{equation}
and the prefactor
\begin{equation}
    |\boldsymbol{W}^J|\approx1-\frac{N_J}{2}
\end{equation}
guarantees that the state is normalized to first order in $N_J\ll1$. 
Although the expansion required $N_J\ll1$, $|\mathring{\beta}|\ll 1$ is the only criteria we need to call the state weakly squeezed; small $|\mathring{\beta}|$ means that we can write Eq.~\eqref{WeakKet} \textit{if} we choose a time window such that $N_J$ is also small. We refer to a partition of weakly squeezed light where Eq.~\eqref{WeakKet} holds as a \textit{single pair window}. For light with a finite pulse length and mean pair number $N\ll1$, the entire pulse can be considered a single pair window. In the CW limit where where $|\beta|\to\infty$ and $N$ diverges, $|\mathring{\beta}|$ will remain finite and we can find a single pair window if $\betac$ is small.

In a single pair window, we can see from Eq.~\eqref{TwoPhotonState} that the state is a superposition of pairs with probability amplitudes $|\boldsymbol{W}^J|T^J_{nm}$, and hence if the supermodes $x$ in the signal and $y$ in the idler are detected, they must be from the pair corresponding to $T^J_{xy}$. The next order in the expansion of Eq.~\eqref{LocalKet} is
\begin{equation}
    \sum_{n,m,j,k}T^J_{nm}T^J_{jk}A^\dagger_nB^\dagger_mA^\dagger_jB^\dagger_k\ket{\vac},
\end{equation}
subject to an appropriate normalization. The detection of signal and idler supermodes $x$ and $y$ could result from the state $T^J_{xy}T^J_{pq}A^\dagger_xB^\dagger_yA^\dagger_pB^\dagger_q\ket{\vac}$, where $x$ and $y$ are from the ``same pair'', but could also result from $T^J_{xq}T^J_{py}A^\dagger_xB^\dagger_qA^\dagger_pB^\dagger_y\ket{\vac}$, where $x$ and $y$ are from ``different pairs''. Photons from the same pair will have a stronger dependence on the properties of $\boldsymbol{T}^J$. For example, consider a joint amplitude like the double Gaussian which is small far from the diagonal, so that $T^J_{nm}$ is small for $n$ far from $m$. Signal and idler photons from the same pair in supermodes $x$ and $y$ have a probability amplitude proportional to $T^J_{xy}$, so it will be unlikely to detect them in supermodes that are far apart in time. If the photons were from different pairs then their probability amplitude is proportional to $T^J_{xq}T^J_{py}$, which can still be large when $x$ is far from $y$. Thus, the presence of multiple pairs can erode behaviour that is prominent in the single-pair regime; we will see this occur with polarization dependent coincidence detection and the Hong-Ou-Mandel effect.


\subsection{Increasing the Whittaker-Shannon Resolution}
\label{Resolution}
Our discussion up to this point has assumed we have chosen a minimal bandlimit $\Omega$ that is just large enough to contain all significant parts of $\gamma(\omega_1,\omega_2)$. However, we could always increase $\Omega$ by an arbitrary amount, and the Whittaker-Shannon decomposition would remain valid. Increasing the bandlimit decreases $\tau$, effectively increasing the temporal resolution of the Whittaker-Shannon decomposition, and in later sections we show many instances where this is useful. But one needs to be careful with the arbitrary nature of $\Omega$ when discussing weak squeezing; $\betac$ is proportional to $\tau$, so by increasing the resolution we could force $\betac\ll1$ for any joint amplitude. For a minimal bandlimit, $\tau$ is on the order of the coherence time \cite{drago2023takingapartsqueezedlight}, and it makes sense to call Eq.~\eqref{WeakKet} the ket in a single pair window since the time window chosen will be at least on the order of $\tau$. On the other hand, if $\tau$ is smaller than the coherence time, Eq.~\eqref{WeakKet} could refer to photons in a time window smaller than that time, and we cannot claim there is at most one photon pair within the coherence time.
Therefore, the quantification of the squeezing strength by $\betac$ requires the use of a bandlimit on the order of the bandwidth.

\subsection{N and M Moments}
So far we have dealt with nondegenerate squeezed states, but it will be useful to outline some results for both degenerate and nondegenerate squeezed states here. A degenerate squeezed ket can can be written as
\begin{equation}
\label{degen ket}
    \ket{\psi}=e^{\frac{\beta}{2}\int d\omega_1d\omega_2\gamma(\omega_1,\omega_2)a^\dagger(\omega_1)a^\dagger(\omega_2)-h.c.}\ket{\vac},
\end{equation}
where the joint spectral amplitude is symmetric in its variables ($\gamma(\omega_2,\omega_1)=\gamma(\omega_1,\omega_2)$). The Whittaker-Shannon decomposition follows analogously to that of nondegenerate squeezed light, and can be seen in full detail in \cite{drago2023takingapartsqueezedlight}.

We define the $\boldsymbol{N}$ and $\boldsymbol{M}$ moments of the Whittaker-Shannon supermodes as
\begin{equation}
    \begin{aligned}
        N^d_{nm}&=\bra{\psi} A^\dagger_nA_m\ket{\psi}\\
        M^d_{nm}&=\bra{\psi}A_nA_m\ket{\psi}
    \end{aligned}
\end{equation}
for degenerate squeezed light, and
\begin{equation}
    \begin{aligned}
        N^a_{nm}&=\bra{\psi} A^\dagger_nA_m\ket{\psi}\\
        N^b_{nm}&=\bra{\psi} B^\dagger_nB_m\ket{\psi}\\
        M^{ab}_{nm}&=\bra{\psi}A_nB_m\ket{\psi}
    \end{aligned}
\end{equation}
for nondegenerate squeezed light. As shown in Appendix \ref{Moments Dix}, they can be written in terms of Whittaker-Shannon matrices:
\begin{equation}
\label{degen_NM}
    \begin{aligned}
        \boldsymbol{N}^d&=\sinh^2\boldsymbol{P}^J\\
        \boldsymbol{M}^d&=(\sinh\boldsymbol{Q}^J)(\cosh\boldsymbol{Q}^J)\boldsymbol{U}^J
    \end{aligned}
\end{equation}
in the degenerate regime, and
\begin{equation}
\label{nondegen_NM}
    \begin{aligned}
        \boldsymbol{N}^a&=(\sinh^2\boldsymbol{Q}^J)^T\\
        \boldsymbol{N}^b&=\sinh^2\boldsymbol{Q}^J\\
        \boldsymbol{M}^{ab}&=(\sinh\boldsymbol{Q}^J)(\cosh\boldsymbol{Q}^J)\boldsymbol{U}^J
    \end{aligned}
\end{equation}
in the nondegenerate regime. Additionally, Eq.~\eqref{nondegen_NM} provides the moments of the state given by Eq.~\eqref{standard_discretization_decomp} \cite{Quesada:22}.

The moments of the CT modes are related to those above by a sum over the Whittaker-Shannon modes. For degenerate squeezed light we have
\begin{equation}
\label{covariance functions}
    \begin{aligned}
        N^d(t,t')&\equiv\bra{\psi}\ol{a}^\dagger(t)\ol{a}(t')\ket{\psi}=\ol{\chi}_n^*(t)N_{nm}^d\ol{\chi}_m(t'),\\
        M^d(t,t')&\equiv\bra{\psi}\ol{a}(t)\ol{a}(t')\ket{\psi}=\ol{\chi}_n(t)M_{nm}^d\ol{\chi}_m(t'),\\
    \end{aligned}
\end{equation}
and for nondegenerate squeezed light,
\begin{equation}
\label{nondegen moments}
    \begin{aligned}
        N^a(t_1,t_2)&\equiv\bra{\psi}\ol{a}^\dagger(t_1)\ol{a}(t_2)\ket{\psi}=\ol{\chi}^*_n(t_1)N^a_{nm}\ol{\chi}_m(t_2),\\
        N^b(t_1,t_2)&\equiv\bra{\psi}\ol{b}^\dagger(t_1)\ol{b}(t_2)\ket{\psi}=\ol{\chi}^*_n(t_1)N^b_{nm}\ol{\chi}_m(t_2),\\
        M^{ab}(t_1,t_2)&\equiv\bra{\psi}\ol{a}(t_1)\ol{b}(t_2)\ket{\psi}=\ol{\chi}_n(t_1)M^{ab}_{nm}\ol{\chi}_m(t_2).
    \end{aligned}
\end{equation}

\section{Homodyne Detection}
\label{Homodyne Sec}
\begin{figure}[h!]
    \centering
    \includegraphics[width=0.45\textwidth]{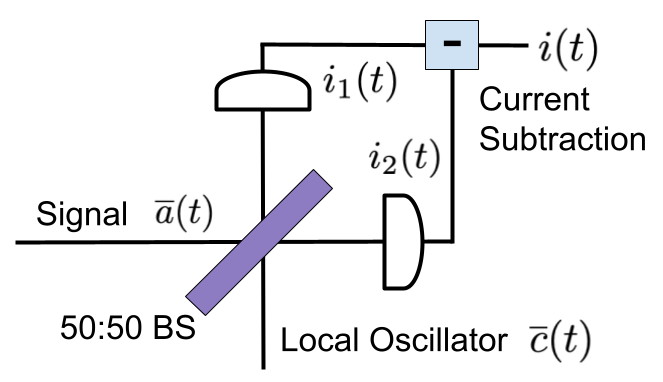}
    \caption{Homodyne Detection Scheme. The signal of interest (for us, a multimode degenerate squeezed state, represented by $\ol{a}(t)$) are mixed on a 50:50 beam splitter with a local oscillator represented by $\ol{c}(t)$. Photodetectors produce currents $i_1(t)$ and $i_2(t)$, and we measure the differential current $i(t)=i_1(t)-i_2(t)$.}
    \label{Homo Scheme}
\end{figure}
To further justify the use of $\betac$ as a quantification of squeezing strength, we investigate the quadrature variance of squeezed light in terms of the Whittaker-Shannon decomposition. Homodyne detection mixes a signal with a strong local oscillator to measure a photocurrent proportional to a quadrature operator of the electromagnetic field \cite{Porto:18,Feng:16}; a schematic is shown in Fig.~\ref{Homo Scheme}. For squeezed light, measurements of the photocurrent variance will fall below the shot-noise limit for certain quadratures, and the minimum variance quantifies the amount of squeezing. In this section we investigate how homodyne detection can be formulated in terms of the Whittaker-Shannon decomposition. We find an expression for the frequency spectrum of the photocurrent variance in the CW limit, and solve for the minimum variance and optimal local oscillator for a measurement of the total homodyne current. For both pulsed and CW homodyne measurement, the minimum variance in dB decreases linearly with $\betac$.

Let us start by considering degenerate multimode squeezed light in the state
\begin{equation}
    \ket{\psi}=e^{\frac{\beta}{2}\int dt_1dt_2\ol{\gamma}(t_1,t_2)\ol{a}^\dagger(t_1)\ol{a}^\dagger(t_2)-h.c.}\ket{\vac}.
\end{equation}

\begin{figure}
    \centering
    \includegraphics[width=0.45\textwidth]{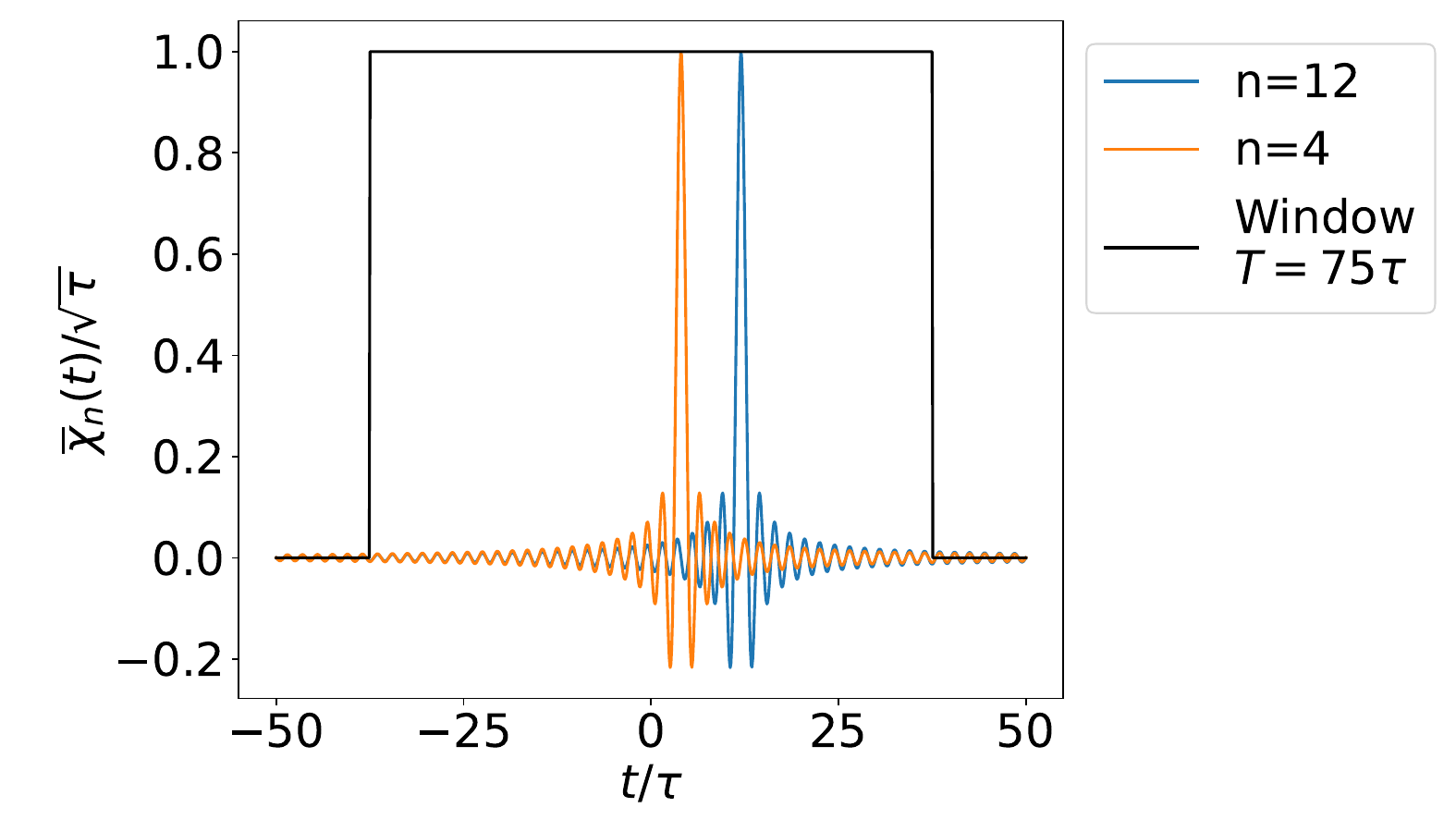}
    \caption{Whittaker-Shannon modes $\overline{\chi}_n(t)$ for $n=4,12$ inside a time window $T\gg\tau$, they are approximately orthogonal inside the window. If $T$ is sufficiently large there will be enough modes inside the window that we can neglect edge effects.}
    \label{LongWindowFig}
\end{figure}

Given sufficiently fast detectors, the photocurrents are proportional to the photon numbers \cite{Patera_2009}, and the difference is represented by the operator
\begin{equation}
    i(t)=\ol{a}^\dagger(t)\ol{c}(t)+\ol{a}(t)\ol{c}^\dagger(t).
\end{equation}
Firstly, we will work in the CW limit where the local oscillator is also CW. Let the local oscillator be in the coherent state $\ket{\eta}$ of $\ol{c}(t)$, where $\eta=|\eta|e^{i\theta}$ for some phase $\theta$ relative to the signal. In the CW limit we can measure quadrature squeezing by analyzing the frequency spectrum of the photocurrent variance (normalized by the local oscillator magnitude $|\eta|^2$) \cite{Feng:16}
\begin{equation}
    \sigma^2_{CW}(\theta,\omega,t)=\frac{1}{|\eta|^2}\int d\tilde{\tau} \langle i(t)i(t+\tilde{\tau})\rangle e^{-i\omega\tilde{\tau}},
\end{equation}
where
\begin{equation}
\label{Current Covariance}
    \begin{aligned}
        \langle i(t)i(t+\tilde{\tau})\rangle&\equiv\bra{\psi}\bra{\eta}i(t)i(t+\tilde{\tau})\ket{\eta}\ket{\psi}.
    \end{aligned}
\end{equation}
Assuming the local oscillator is much stronger than the signal, we find
\begin{widetext}
\begin{equation}
    \langle i(t)i(t+\tilde{\tau})\rangle=|\eta|^2\big(\delta(\tilde{\tau})+N^d(t,t+\tilde{\tau})+N^d(t+\tilde{\tau},t)+e^{2i\theta}M^d(t,t+\tilde{\tau})+e^{-2i\theta}\big(M^d(t+\tilde{\tau},t)\big)^*\big).
\end{equation}
\end{widetext}
Although we should be in a stationary state because we are working in the CW limit, the Whittaker-Shannon decomposition introduced time dependence based on the Whittaker-Shannon modes, so we take an average over some time window $ J=[t_0-T/2,t_0+T/2]$:
\begin{equation}
    \sigma^2_{CW}(\theta,\omega)=\frac{1}{T}\int_Jdt\sigma^2_{CW}(\theta,\omega,t).
\end{equation}
If the time window is much larger than the Whittaker-Shannon timescale ($T\gg\tau$), then the Whittaker-Shannon modes $\overline{\chi}_n(t)$ will be approximately orthonormal within $J$ (Fig.~\ref{LongWindowFig}). In terms of the matrices $\boldsymbol{Q}^J$ etc. restricted to the time window, the variance is
\begin{widetext}
\begin{equation}
\label{V CW}
    \sigma^2_{CW}(\theta,\omega)=1+\frac{\tau}{T}\Tr\left[\boldsymbol{E}(\omega)(\sinh^2\boldsymbol{Q}^J)+\boldsymbol{E}^T(\omega)(\sinh^2\boldsymbol{Q}^J)+2\RE\{e^{2i\theta}\boldsymbol{E}^T(\omega)(\sinh\boldsymbol{Q}^J)(\cosh\boldsymbol{Q}^J)\boldsymbol{U}^J\}\right],
\end{equation}
\end{widetext}
where $E_{nm}(\omega)=e^{i(n-m)\omega\tau}$ (Appendix \ref{CW Dix}). Observe that $\sigma^2_{CW}(\theta,\omega)$ is normalized so that the variance of the vacuum state is 1, and should not depend on $t_0$ or $T$ as long as $T$ is large enough. For a joint spectral amplitude that extends slightly outside the bandlimit, Eq.~\eqref{V CW} is inaccurate when $\omega$ approaches $\pm\frac{\Omega}{2}$ since the frequencies outside of the bandlimit are not included. Increasing the bandlimit to rectify this issue is equivalent to increasing the Whittaker-Shannon resolution (Section \ref{Resolution}).

\begin{figure}
    \centering
    \includegraphics[width=0.45\textwidth]{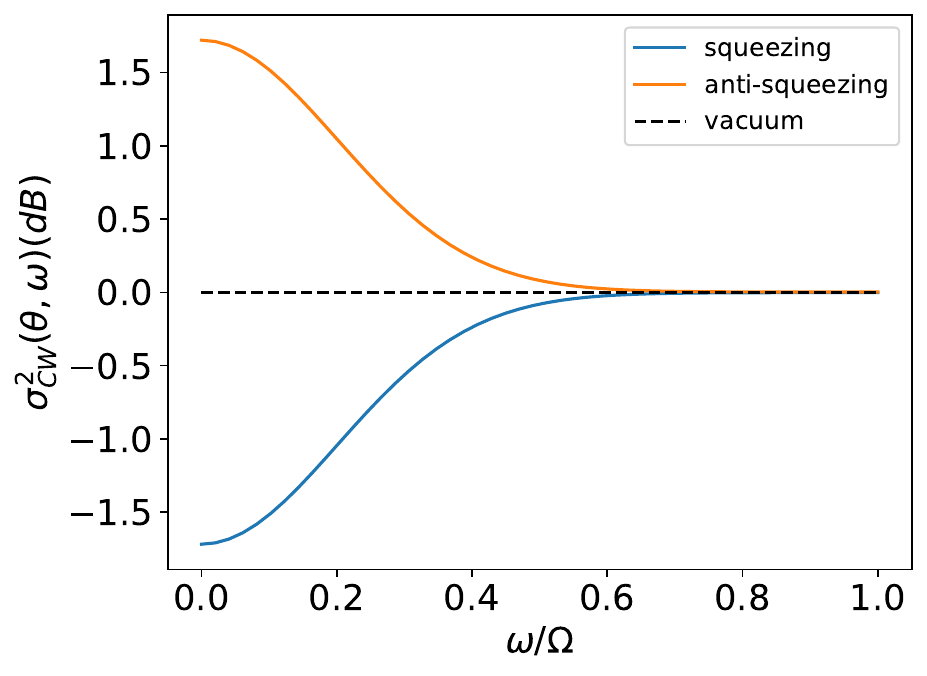}
    \caption{Variance of spectral squeezing ($\theta=\pi/2$) and anti-squeezing ($\theta=0$) vs. $\omega$ for the double Gaussian joint amplitude in the CW limit with $\betac=0.1$, in a time window of size $d_J=60$ centered at $t=0$. The most squeezing is obtained for $\omega=0$ at the center of the joint spectral amplitude, and squeezing is reduced as $\omega$ gets farther from the center. To maintain accuracy for $\omega$ close to $\Omega/2$ and show where the squeezing goes to zero, this calculation used an increased bandlimit of $\Omega'=2\Omega$.}
    \label{CW Homo Fig}
\end{figure}
In Fig.~\ref{CW Homo Fig} we plot the minimum (squeezing) and maximum (anti-squeezing) variance for a double Gaussian joint amplitude in the CW limit vs. $\omega$, and observe that the variance is modulated more when $\omega$ is closer to zero.
Since $\boldsymbol{U}^J$ is the identity matrix for the double Gaussian joint amplitude, the minimum variance occurs at $\theta=\pi/2$, and the maximum at $\theta=0$.

\begin{figure}
    \centering
    \includegraphics[width=0.5\textwidth]{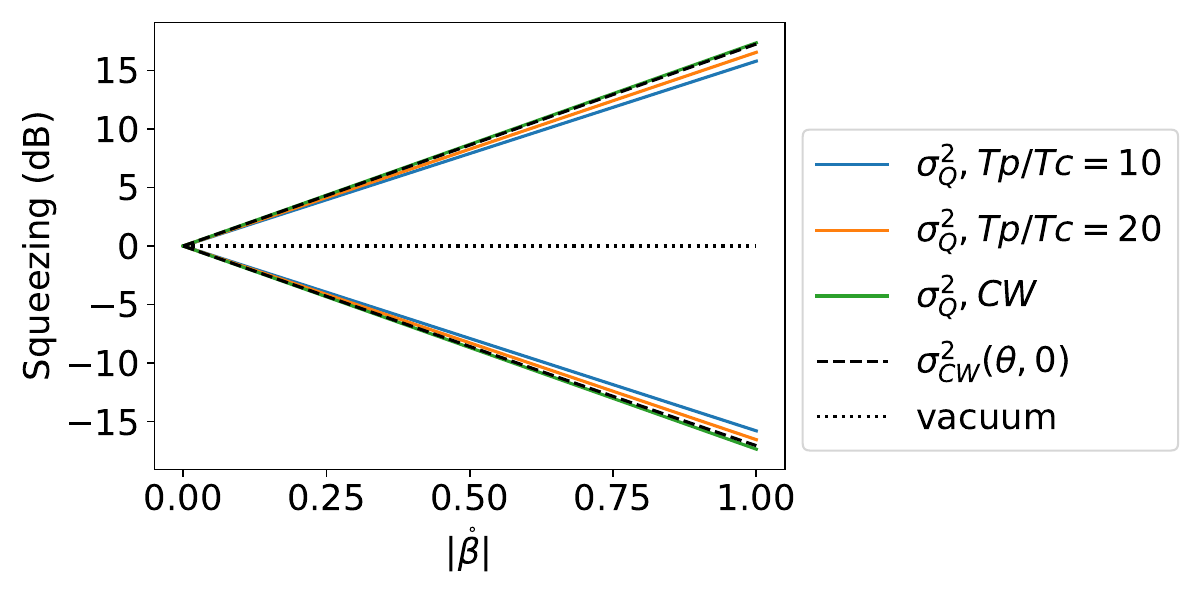}
    \caption{Squeezing (upward slopes) and anti-squeezing (downward slopes) vs. $\betac$ of the double Gaussian in a time window of size $d_J=60$ centered at $t=0$. The squeezing of the total homodyne charge measurement increases as the ratio $T_p/T_c$ gets larger. We also plot the squeezing of a spectral analysis homodyne measurement at $\omega=0$ in the CW limit; it is not quite as strong as for the total charge measurement in the CW limit, aligning with the fact that spectrum analysis is not necessarily the optimal homodyne measurement \cite{Shapiro:97}. For all scenarios, the squeezing (in dB) depends linearly on $\betac$.}
    \label{Min Homo Fig}
\end{figure}

More generally, we consider measurements of the total ``homodyne charge'' within the time window $J$ \cite{Shapiro:97}
\begin{equation}
    Q=\int_Jdt\;i(t).
\end{equation}
If we define the normalized strong local oscillator
\begin{equation}
    \xi(t)=\frac{\bra{\psi}c(t)\ket{\psi}}{\sqrt{N_c}},
\end{equation}
where $N_c$ is the photon number expectation value of the local oscillator over the time window, then the normalized variance in a measurement of $Q$ is
\begin{equation}
    \sigma^2_Q=1+2\int_Jdtdt'\boldsymbol{\xi}^T(t)\boldsymbol{K}(t,t')\boldsymbol{\xi}(t'),
\end{equation}
where
\begin{equation}
    \boldsymbol{K}(t,t')=\begin{pmatrix}
        N^d_R(t,t')+M^d_R(t,t') & N^d_I(t,t')+M^d_I(t,t')\\
        M^d_I(t,t')-N^d_I(t,t') & N^d_R(t,t')-M^d_R(t,t')
    \end{pmatrix},
\end{equation}
and
\begin{equation}
    \boldsymbol{\xi}(t)=\begin{pmatrix}
        \xi_R(t)\\
        \xi_I(t)
    \end{pmatrix},
\end{equation}
with the subscripts $R$ and $I$ denoting real and imaginary parts, respectively \cite{Shapiro:97}. The minimum variance is
\begin{equation}
    \min(\sigma^2_Q)=1+2\lambda_{min},
\end{equation}
where $\lambda_{min}$ is the smallest eigenvalue of the Fredholm integral equation
\begin{equation}
\label{Fredholm Eqn}
    \int_Jdt'\boldsymbol{K}(t,t')\boldsymbol{\phi}_n(t')=\lambda_n
    \boldsymbol{\phi}_n(t),
\end{equation}
given the set of real-valued vector eigenfunctions $\boldsymbol{\phi}_n$. The minimum variance is achieved when $\boldsymbol{\xi}(t)$ is the eigenfunction $\boldsymbol{\phi}_{min}$ corresponding to $\lambda_{min}$ \cite{Shapiro:97}. The same applies to the maximum variance and the largest eigenvalue $\lambda_{max}$. Since the Whittaker-Shannon modes are approximately complete, we let $\boldsymbol{\phi}_n(t)=\sum_j\boldsymbol{\varphi}^n_j \ol{\chi}_j(t)$, and this along with Eq.~\eqref{covariance functions} and assuming that $T\gg\tau$ simplifies the Fredholm equation to the matrix eigenvalue equation
\begin{equation}
    \label{Matrix Eig Eqn}
    \begin{aligned}
        \boldsymbol{K}\boldsymbol{\Phi}_n=\lambda_n\boldsymbol{\Phi}_n,
    \end{aligned}
\end{equation}
where $\boldsymbol{K}$ is the block matrix
\begin{equation}
    \boldsymbol{K}=\begin{pmatrix}
           \boldsymbol{N}^d_R+\boldsymbol{M}^d_R & \boldsymbol{N}^d_I+\boldsymbol{M}^d_I\\
           \boldsymbol{M}^d_I-\boldsymbol{N}^d_I & \boldsymbol{N}^d_R-\boldsymbol{M}^d_R
       \end{pmatrix},
\end{equation}
and $\boldsymbol{\Phi}_n$ is the block vector
\begin{equation}
    \boldsymbol{\Phi}_n=\begin{pmatrix}
        \boldsymbol{\varphi}^n_R\\
        \boldsymbol{\varphi}^n_I
    \end{pmatrix},
\end{equation}
with $\boldsymbol{\varphi}^n$ the vector with components $\varphi^n_j$. The local oscillator that results in minimum variance is again that which corresponds to the eigenvector $\boldsymbol{\phi}_{min}$ that has the minium eigenvalue. In Fig.~\ref{Min Homo Fig} we plot the squeezing and anti-squeezing against $\betac$ for both $\sigma^2_Q$ and $\sigma^2_{CW}$. In all cases, the minimum variance (in dB) decreases linearly with $\betac$, illustrating the close relationship between $\betac$ and the typical indicator of squeezing strength \cite{Strong_Optomechanical_Squeezing,Intense_Squeezing,10dB_strength}.

\section{Polarization Dependent Coincidence Detection}
\label{Coincidence Sec}

\begin{figure}
    \centering
    \includegraphics[width=0.45\textwidth]{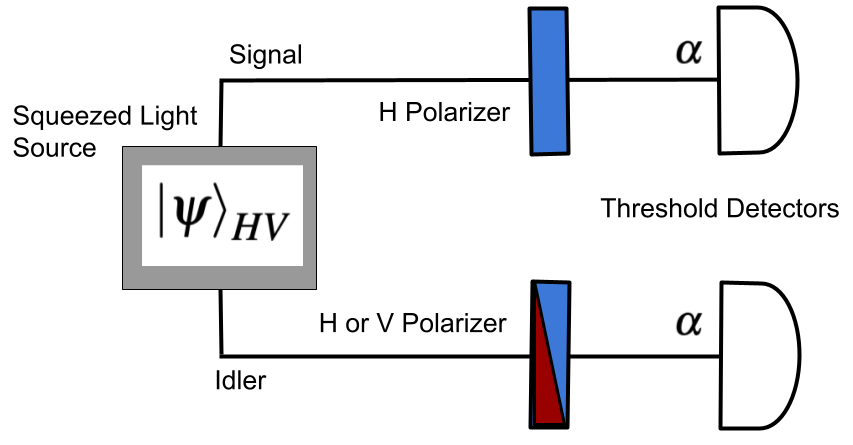}
    \caption{Coincidence detection scheme. The signal modes pass through a polarizer fixed at $H$, and the polarizer applied to the idler modes is either $H$ or $V$. When the idler polarizer is $H$ ($V$), we imagine finding the probability $\mathcal{P}_{HH}$ ($\mathcal{P}_{HV}$) for both detectors to click.}
    \label{Scheme}
\end{figure}

As a second example, we use the Whittaker-Shannon decomposition to characterize the coincidence detection probabilities of a two-polarization nondegenerate multimode squeezed state, with $H$ ($V$) polarized signal and idler modes labeled by $a_H(\omega_1)$ ($a_V(\omega_1)$) and $b_H(\omega_2)$ ($b_V(\omega_2)$), respectively. Following Takesue \cite{TAKESUE2010276}, we take the Hamiltonian to be
\begin{equation}
    \hat{H}(t)=\hat{H}_H(t)+\hat{H}_V(t),
\end{equation}
where the time evolution of $\hat{H}_H(t)$ ($\hat{H}_V(t)$) results in a nondegenerate squeezed vacuum state in the $H$ ($V$) polarization
\begin{equation}\begin{aligned}
    \ket{\psi_H}=e^{\beta\int dt_1dt_2\overline{\gamma}(t_1,t_2)\overline{a}_H^\dagger(t_1)\overline{b}_H^\dagger(t_2)-h.c.}\ket{\vac_H},
\end{aligned}\end{equation}
and similarly for $\ket{\psi_V}$. Since the total Hamiltonian is a sum of those for each polarization, the resulting state is separable as
\begin{equation}\begin{aligned}
    \ket{\psi_{HV}}&=\ket{\psi_H}\otimes\ket{\psi_V}\\
    \ket{\psi_{HV}}&=e^{\beta_{nm}{A_n^H}^\dagger{B_m^H}^\dagger-h.c.}\ket{\vac_H}\\
    &\quad\otimes e^{\beta_{nm}{A_n^V}^\dagger{B_m^V}^\dagger-h.c.}\ket{\vac_V},
\end{aligned}\end{equation}
where we took the Whittaker-Shannon decomposition in each polarization using definitions of $\beta_{nm}$, ${A_n^H}^\dagger$, etc. analogous to those in Section \ref{Formalism Section}. We have assumed that both polarizations share the same joint amplitude, but the results can be easily generalized to the scenario where they are different. The detection scheme shown in Fig.~\ref{Scheme} allows us to measure the coincidence detection probability between $H$ photons in the signal and idler ranges $\mathcal{P}_{HH}$ and the coincidence detection probability between $H$ signal photons and $V$ idler photons $\mathcal{P}_{HV}$. Since the $H$ and $V$ squeezed states are uncorrelated, comparing $\mathcal{P}_{HH}$ and $\mathcal{P}_{HV}$ tells us the degree to which coincidence counts are due to correlations between signal and idler photons.

Imagine turning on the detectors for a long time $T\gg\tau$, we consider a coincidence to occur if both detectors fire at least once within the time window, regardless of the time between each detector firing. If we take the detection probability of $x$ photons incident on a threshold detector of efficiency $\alpha$ to be $D_x\equiv1-(1-\alpha)^x$ \cite{TAKESUE2010276}, then we can find the coincidence detection probabilities by summing over the probability of projecting onto each possible photon number combination, weighted by their detection probabilities. Let $\ket{\vac}_{a_H}^{ J}$ be the vacuum state corresponding to the $H$ polarized signal modes within the time window $ J=[t_0-T/2,t_0+T/2]$, and $\mathbb{I}_{a_H}^{\notJ}$ be the identity operator for the $H$ polarized signal modes outside $ J$, then $V_{a_H}^ J\equiv\ket{\vac}_{a_H}^{ J}\bra{\vac}_{a_H}^{ J}\otimes\mathbb{I}_{a_H}^{\notJ}$ is the operator that projects onto the vacuum state for modes within the time window, and onto the identity for modes outside it. The projector $P^{a_H}_{J,s}$ for $s$ signal photons within $ J$ can be written as (Appendix \ref{Projector Dix})
\begin{equation}
\label{projector}
    P^{a_H}_{J,s}=\frac{1}{s!}\int_J dt_1\hdots dt_s\ol{a}_H^\dagger(t_1)\hdots \ol{a}_H^\dagger(t_s)V_{a_H}^ J\ol{a}_H(t_1)\hdots \ol{a}_H(t_s),
\end{equation}
and similarly for $P^{b_H}_{J,s}$ and $P^{b_V}_{J,s}$. We can then write the coincidence probabilities as
\begin{equation}\begin{aligned}
\label{PHH and PHV}
    \mathcal{P}_{HH}&=\sum_{s=1}^\infty D_s^2\bra{\psi_H}P^{a_H}_{J,s}P^{b_H}_{J,s}\ket{\psi_H},\\ \mathcal{P}_{HV}&=\sum_{s_a,s_b=1}^\infty D_{s_a}D_{s_b}\bra{\psi_H}P^{a_H}_{J,s_a}\ket{\psi_H}\\
    &\quad\quad\quad\quad\quad\quad\times\bra{\psi_V}P^{b_V}_{J,s_b}\ket{\psi_V}.
\end{aligned}\end{equation}
The expression for $\mathcal{P}_{HH}$ has a single sum since there must be the same number of signal and idler photons. 

Since $T\gg\tau$, the Whittaker-Shannon modes $\overline{\chi}_n(t)$ will be approximately orthonormal within the time window and we can neglect the outside modes (Fig.~\ref{LongWindowFig}), allowing us to approximate
\begin{equation}
    P^{a_H}_{J,s}\approx P^{A^H}_{J,s},
\end{equation}
where
\begin{equation}
    P^{A^H}_{J,s}=\frac{1}{s!}\sum^J_{n_1,\hdots,n_s}A^{H\dagger}_{n_1}\hdots A^{H\dagger}_{n_s} V_{A^H}^JA^H_{n_1}\hdots A^H_{n_s}.
\end{equation}
The superscript $J$ on the sum indicates that each sum over $n_j$ ranges across the indices for which $n_j\tau$ is inside the time window. Analogously to $V_{a_H}^ J$, $V_{A^H}^J=\ket{\vac}_{A^H}^J\bra{\vac}_{A_H}^J\otimes\mathbb{I}_{A_H}^{\notJ}$ projects on to the vacuum for the Whittaker-Shannon modes with $n\tau$ inside the time window, and onto the identity otherwise. Essentially we have approximated the set of CT modes in the time window as the set $J$ of Whittaker-Shannon modes that are centered inside the window. To count coincidences in a time window where $T\gg\tau$ does not apply we could increase the Whittaker-Shannon resolution.

\begin{figure}
    \centering
    \includegraphics[width=0.5\textwidth]{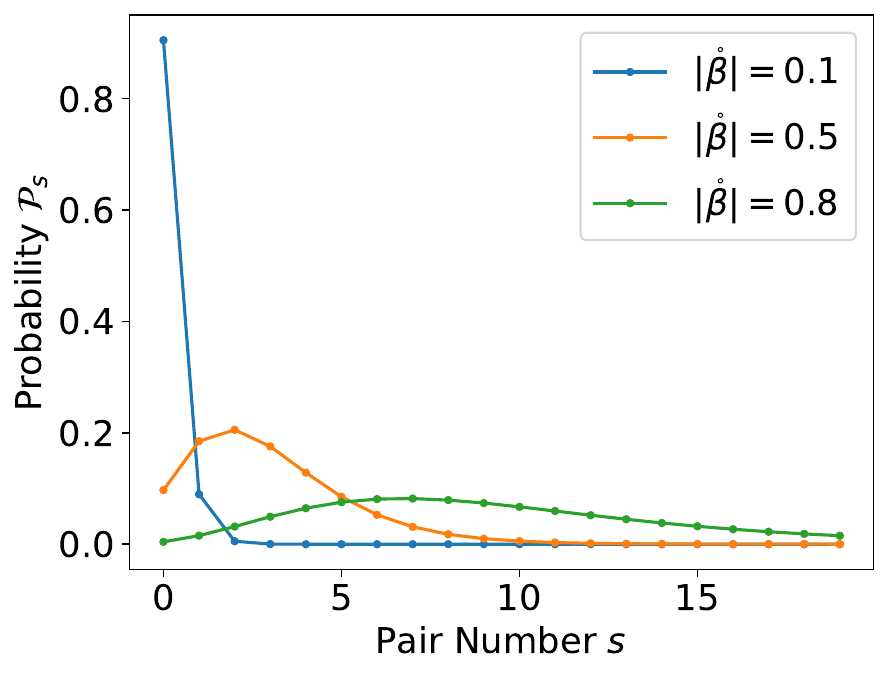}
    \caption{Pair number probabilities $\mathcal{P}_s$ vs. pair number $s$ for the double Gaussian joint amplitude with $T_p/T_c=10$. For small $|\mathring{\beta}|$ we are in the weakly squeezed regime and are most likely to find zero or one pairs. As $|\mathring{\beta}|$ increases higher numbers of pairs are more likely, but the probability still approaches zero for large enough $s$.}
    \label{Num_Prob_Plot}
\end{figure}

Now the coincidence detection probabilities can be rewritten in terms of the Whittaker-Shannon projectors $P^{A^H}_{J,s}$, $P^{B^H}_{J,s}$, and $P^{B^V}_{J,s}$:
\begin{equation}\begin{aligned}
\label{LargeTProbsWS}
    \mathcal{P}_{HH}&=\sum_{s=1}^\infty D_s^2\bra{\psi_H}P^{A^H}_{J,s}P^{B^H}_{J,s}\ket{\psi_H},\\ \mathcal{P}_{HV}&=\sum_{s_a,s_b=1}^\infty D_{s_a}D_{s_b}\bra{\psi_H}P^{A^H}_{J,s_a}\ket{\psi_H}\bra{\psi_V}P^{B^V}_{J,s_b}\ket{\psi_V}.
\end{aligned}\end{equation}
Since signal and idler photons are created in pairs, the probabilities to find a signal photon and both a signal and idler photon are equal, so we define the pair probability $\mathcal{P
}^H_s\equiv\bra{\psi_H}P^{A^H}_{J,s}\ket{\psi_H}=\bra{\psi_H}P^{A^H}_{J,s}P^{B^H}_{J,s}\ket{\psi_H}$, and similarly for $V$. Since we assumed that the squeezed light in each polarization has the same joint amplitude, $\mathcal{P}_s^V=\mathcal{P}_s^H\equiv\mathcal{P}_s$. In Appendix \ref{LargeTDix} we show that
\begin{equation}
\label{Num_Prob}
    \begin{aligned}
        \mathcal{P}_s=\sum_{\{q_n\}\vdash s}\frac{| \boldsymbol{W}^J|^2}{1^{q_1}(q_1!)\hdots s^{q_s}(q_s!)}\prod_{u=1}^s\Tr((\tanh^2 \boldsymbol{Q}^J)^u)^{q_n},
    \end{aligned}
\end{equation}
where $\{q_n\}\vdash s$ is the integer partition of $s$ for which $u$ appears $q_u$ times and we sum over all possible integer partitions. In Fig.~\ref{Num_Prob_Plot} we plot $\mathcal{P}_s$ for different values of $|\mathring{\beta}|$ and confirm that higher pair numbers are much more likely with large squeezing strength. If the $H$ and $V$ polarizations had different joint amplitudes, $ \boldsymbol{W}^J$ and $ \boldsymbol{Q}^J$ would depend on the polarization; the other results can be generalized in a similar manner.

Since $\mathcal{P}^H_s$ and $\mathcal{P}^V_s$ become negligible at large enough $s$, we can find the coincidence detection probability up to some desired precision by computing a finite number of terms in Eq.~\eqref{LargeTProbsWS}. However, we can find analytic expressions for $\mathcal{P}_{HH}$ and $\mathcal{P}_{HV}$ in a few limits of detection efficiency and squeezing strength.

First, we consider two special cases of detection efficiency, \textit{case 1} where we have perfect detection efficiency ($\alpha$=1), and \textit{case 2} where we have  small detection efficiency. The expressions we provide for case 1 and case 2 are valid for any squeezing strength, and are derived in Appendix \ref{LargeTDix}. In case 1 the coincidence detection probabilities are given by
\begin{equation}
\begin{aligned}
\label{perfect efficiency P}
    \mathcal{P}_{HH}&=1-| \boldsymbol{W}^J|^2, & \mathcal{P}_{HV}&=(1-| \boldsymbol{W}^J|^2)^2.
\end{aligned}    
\end{equation}
In case 2 we can expand in $\alpha$ to find $D_s\approx \alpha s+\mathcal{O}(\alpha^2s^2)$, but this is not necessarily valid for $\alpha\ll1$ since it requires $\alpha s\ll1$ and we sum over all photon numbers $s$. However, we can set sufficient conditions for the small $\alpha$ expansion depending on when the photon number probabilities drop off; for $\mathcal{P}_{HH}$ we need $\alpha s\ll1$ when $\mathcal{P}_s$ is significant, and for $\mathcal{P}_{HV}$ we need $\alpha s_a\ll1$ and $\alpha s_b\ll1$ when $\mathcal{P}_{s_a}\mathcal{P}_{s_b}$ is significant. This makes the approximation more suited to smaller $|\mathring{\beta}|$, but it can be valid for any $\betac$ provided that $\alpha$ is small enough. 
If the approximation holds then we can write the coincidence probabilities for small detection efficiency as
\begin{equation}\begin{aligned}
\label{small efficiency P}  \mathcal{P}_{HH}&=\alpha^2\bigg(N_J+N_J^2+\Tr(\sinh^4 \boldsymbol{Q}^J)\bigg),
\\\mathcal{P}_{HV}&=\alpha^2N_J^2,
\end{aligned}\end{equation}
where $N_J=\Tr(\sinh^2 \boldsymbol{Q}^J)$ is the average number of photon pairs in the time window for one of the polarizations. In a model with two modes, Eq.~\eqref{small efficiency P} reduces to those given by Takesue \cite{TAKESUE2010276} (Appendix \ref{Takesue_Equivalence_Dix}).

We can also simplify the general expressions Eq.~\eqref{LargeTProbsWS} for coincidence probabilities for a single pair window ($|\mathring{\beta}|\ll1$ and $N_J\ll1$), so that Eq.~\eqref{WeakKet} applies. Expanding Eq.~\eqref{LargeTProbsWS} up to order $N_J^2$ gives 
\begin{equation}\begin{aligned}
\label{WeakCoincidenceProbs}
    \mathcal{P}_{HH}&=D_1^2N_J+\bigg(\frac{D_2^2}{2}-D_1^2\bigg)N_J^2, &
    \mathcal{P}_{HV}&=D_1^2N_J^2.
\end{aligned}\end{equation}
Unlike special cases 1 and 2 of detection efficiency, these expressions are valid for any value of $\alpha$. A coincidence between $H$ and $V$ requires the small chance of detecting a photon to occur independently in both polarizations, and so is proportional to $N_J^2$. But $\mathcal{P}_{HH}$ has a term proportional to $N_J$ because a coincidence can be detected from just one photon pair. Since $N_J\ll1$, $\mathcal{P}_{HH}$ will be much larger than $\mathcal{P}_{HV}$ in a single pair window. Other investigations into the effects of multiple photon pairs (including Takesue's) apply a heuristic where the number of temporal modes in a long pulse is large enough that there will never be two or more pairs in the same temporal mode \cite{TAKESUE2010276,Zhong_QKD_model}. The Whittaker-Shannon formalism provides a more rigorous description of this scenario, and Eq.~\eqref{WeakCoincidenceProbs} agrees with Takesue \cite{TAKESUE2010276} (Appendix \ref{Takesue_Equivalence_Dix}).



\begin{figure}
    \centering
    \includegraphics[width=0.45\textwidth]{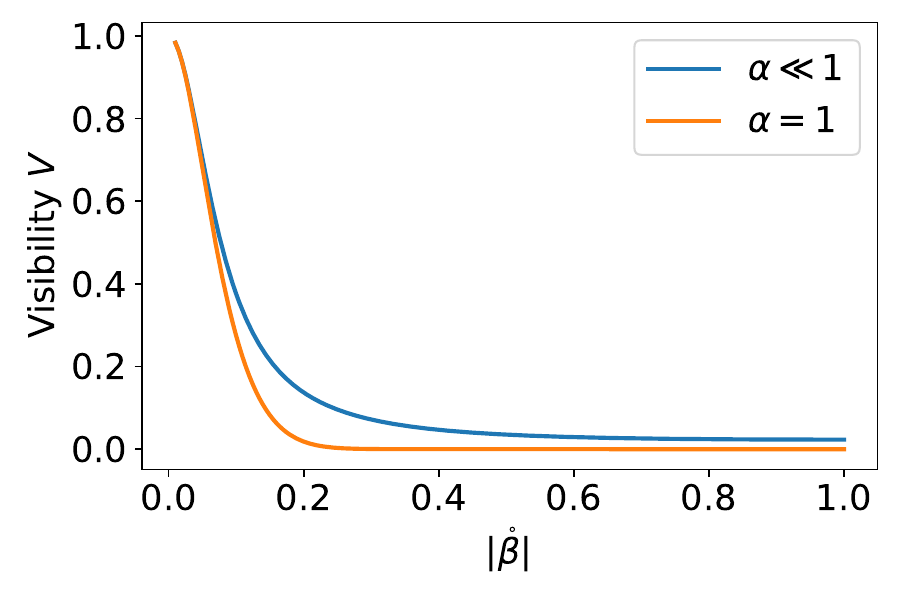}
    \caption{Visibility vs. $|\mathring{\beta}|$ for case 1 (orange) and case 2 (blue) for a double Gaussian joint amplitude in the CW limit, in a time window of size $d_J=60$ centered at $t=0$. The visibility is smaller as $|\mathring{\beta}|$ increases because as there are more photon pairs there is a greater contribution to the coincidence probability from separate pairs.}
    \label{Coincidence_Visibility_Fig}
\end{figure}

\begin{figure}
    \centering
    \includegraphics[width=0.45\textwidth]{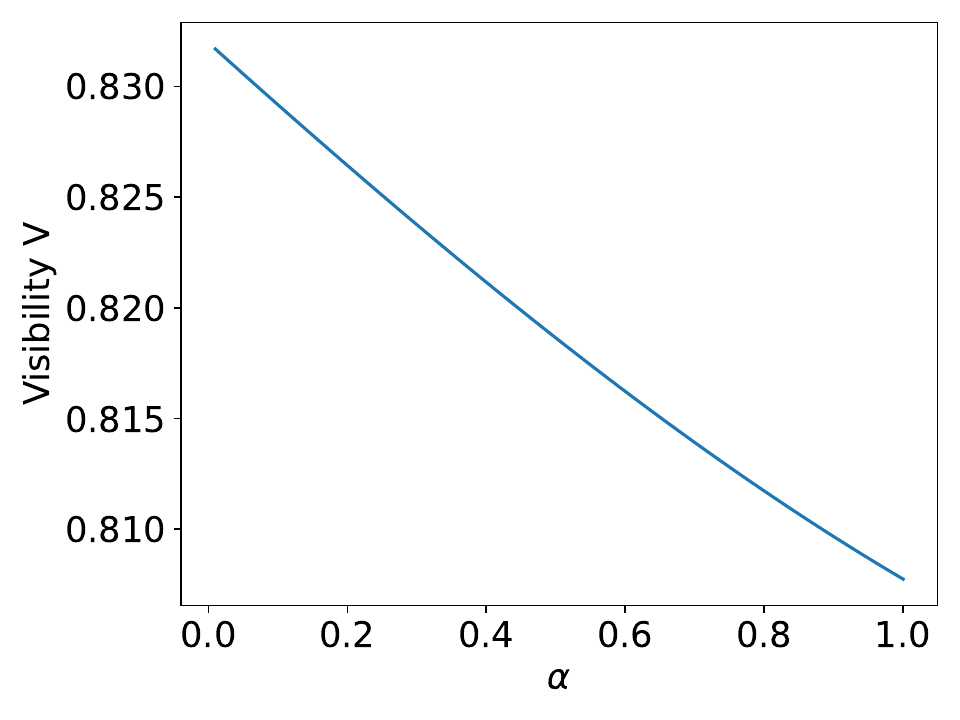}
    \caption{Visibility vs. detection efficiency for a double Gaussian joint amplitude in a single pair window ($\betac=0.1$ and $T_p/T_c=10$). The visibility decreases slightly as detection efficiency rises, matching Fig.~\ref{Coincidence_Visibility_Fig} where small detection efficiency shows improved visibility. }
    \label{Weak Visibility Fig}
\end{figure}

If the angle of the rotatable polarizer were changed, the maximum coincidence probability would be $\mathcal{P}_{HH}$, since the detected photons are most correlated when both detectors see the same polarization; and the minimum coincidence probability would be $\mathcal{P}_{HV}$, since there is no correlation between the $H$ and $V$ states. Therefore, the visibility
\begin{equation}
    V=\frac{\mathcal{P}_{HH}-\mathcal{P}_{HV}}{\mathcal{P}_{HH}+\mathcal{P}_{HV}}
\end{equation}
quantifies how the coincidence probability depends on the correlations between signal and idler. When squeezing is weak and there is only a small probability to detect a pair, $\mathcal{P}_{HH}$ is much larger than $\mathcal{P}_{HV}$ since a coincidence between $H$ and $V$ polarizations requires two independent unlikely events. As squeezing becomes larger, both probabilities increase, but there is less of a difference between $\mathcal{P}_{HH}$ and $\mathcal{P}_{HV}$ since coincidences between separate pairs make up more of the total contribution to $\mathcal{P}_{HH}$. As seen in Fig.~\ref{Coincidence_Visibility_Fig}, the visibility approaches zero for large squeezing in case 1; a coincidence is very likely for either idler polarization since there are so many pairs. In case 2, there remains a nonzero visibility for large squeezing because the detection efficiency is low enough that a coincidence detection is not overwhelmingly likely.
We also plot the visibility for weakly squeezed light in a single pair window against detection efficiency $\alpha$ in Fig.~\ref{Weak Visibility Fig}. Higher detection efficiency increases the chance of detecting photons from uncorrelated pairs of different polarizations, reducing the visibility even in a single pair regime.

\section{Hong-Ou-Mandel Effect}
\label{HOM Sec}

Finally, we consider the Hong-Ou-Mandel effect, which occurs when indistinguishable photons are incident on a beam splitter and destructive interference occurs in one of the output ports. It is often demonstrated with a varying time delay on identical input states; when the time delay is zero they are indistinguishable and the probability to detect light in both output modes approaches zero \cite{DragoHOM}. In this section we apply the Whittaker-Shannon decomposition to the HOM effect and use the ideas we have built about squeezing strength to explain how the HOM effect differs for weakly and strongly squeezed light.

\begin{figure}
    \centering
    \includegraphics[width=0.5\textwidth]{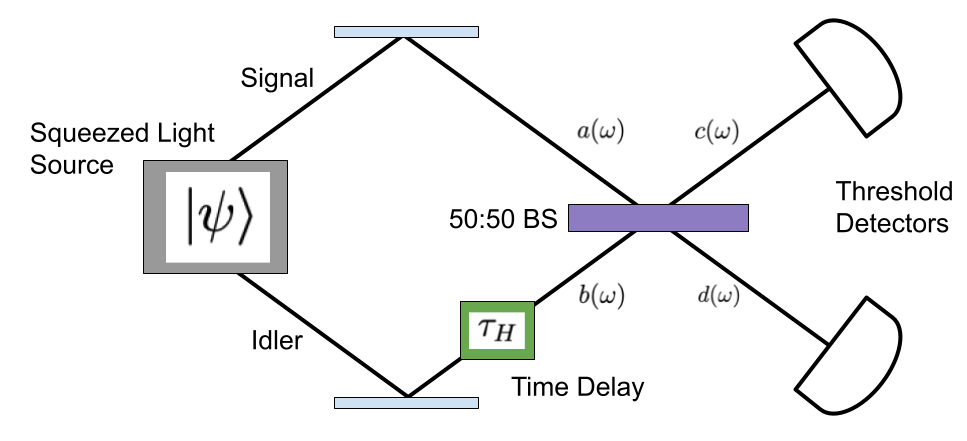}
    \caption{Hong-Ou-Mandel scheme. The idler modes undergo a time delay $\tau_H$ before they are mixed with the signal modes on a 50:50 beam splitter. We consider the probability for both threshold detectors to click, depending on the time delay.}
    \label{HOM Scheme}
\end{figure}
Consider the scheme shown in Fig.~\ref{HOM Scheme} where we induce a time delay $\tau_H$ on the idler modes and then mix them with the signal modes through a 50:50 beam splitter before they are incident on threshold detectors. Our input is a multimode squeezed state where the signal and idler modes share the same center frequency; however, we still label them with separate operators $\ol{a}(t)$ and $\ol{b}(t)$ since they exist in spatially separated channels. This can be written as a nondegenerate squeezed ket
\begin{equation}
    \ket{\psi}=|\boldsymbol{W}|e^{T_{nm}A^\dagger_nB^\dagger_m}\ket{\vac};
\end{equation}
we have used the disentangled form given by Eqs.~\eqref{Disentangling Matrices}\eqref{LocalKet}, but kept the full matrix $\boldsymbol{\beta}$ instead of a partition $\boldsymbol{\beta}^J$.

The time delay transforms $\ol{b}(t)\to\ol{b}(t-\tau_H)$, which shifts the Whitaker-Shannon mode operators to 
\begin{equation}\begin{aligned}
    B_m^\dagger\to\int dt\ol{\chi}_m(t)\ol{b}^\dagger(t-\tau_H)=\int dt \ol{\chi}_m(t+\tau_H)\ol{b}^\dagger(t).
\end{aligned}\end{equation}
If $\tau_H$ is a multiple of the Whittaker-Shannon timescale $\tau$ so that $\tau_H=q\tau$ for some integer $q$, then $\ol{\chi}_m(t+\tau_H)=\ol{\chi}_{(m-q)}(t)$, therefore,
\begin{equation}
    B_m^\dagger\to B_{(m-q)}^\dagger.
\end{equation}
The beam splitter causes the transformation
\begin{equation}
    \begin{aligned}
        \ol{a}(t)&\to\frac{1}{\sqrt{2}}\left(\ol{c}(t)+\ol{d}(t)\right),\\
        \ol{b}(t)&\to\frac{1}{\sqrt{2}}\left(\ol{c}(t)-\ol{d}(t)\right),
    \end{aligned}
\end{equation}
where $\ol{c}(t)$ and $\ol{d}(t)$ are the annihilation operators for the modes associated with the beam splitter outputs, and this transforms the Whittaker-Shannon operators as  
\begin{equation}
    \begin{aligned}
        A^\dagger_n&\to\frac{1}{\sqrt{2}}\left(C^\dagger_n+D^\dagger_n\right)\\
        B^\dagger_m&\to\frac{1}{\sqrt{2}}\left(C^\dagger_m-D^\dagger_m\right),
    \end{aligned}
\end{equation}
where $C^\dagger_n$ and $D^\dagger_n$ are the natural extensions of the Whittaker-Shannon supermode operators to the beam splitter outputs:
\begin{equation}
    \begin{aligned}
        C^\dagger_n&=\int dt\ol{\chi}_n(t)\ol{c}^\dagger(t), & D^\dagger_n&=\int dt\ol{\chi}_n(t)\ol{d}^\dagger(t).
    \end{aligned}
\end{equation}
By defining the shifted matrix $\mathring{T}_{nm}\equiv T_{n,(m+q)}$, we can write the state in the time window $J$ after the time delay and beam splitter as
\begin{equation}
    \ket{\psi_{HOM}(\tau_H)}^J=|\boldsymbol{W}|e^{\frac{1}{2}\mathring{T}^J_{nm}(C^\dagger_n+D^\dagger_n)(C^\dagger_m-D^\dagger_m)}\ket{\vac},
\end{equation}
where $\boldsymbol{\mathring{T}}^J$ is equal to $\boldsymbol{\mathring{T}}$ for indices inside $J$, and zero for indices outside of it. The matrix partition must be taken after applying the time delay, since it changes which Whittaker-Shannon modes are included in $J$.

As in the previous section, we consider a coincidence count to occur when both detectors register a click at least once. We show in Appendix \ref{HOMDix} that with perfect detection efficiency the coincidence probability is
\begin{equation}\begin{aligned}
\label{HOMProbStart}
    &\mathcal{P}_{HOM}(\tau_H)=1+|\boldsymbol{W}^J|^2\left(1-2|\boldsymbol{I}^J-(\boldsymbol{\lambda}^J)^\dagger\boldsymbol{\lambda}^J|^{-\frac{1}{2}}\right),
\end{aligned}\end{equation}
where $\boldsymbol{\lambda}^J\equiv\frac{1}{2}(\boldsymbol{\mathring{T}}^J+(\boldsymbol{\mathring{T}}^J)^T)$ (the symmetrization of $\boldsymbol{\mathring{T}}^J$) and $\boldsymbol{I}^J$ is the identity matrix of appropriate dimension. Although this calculation only works when $\tau_H$ is an integer multiple of $\tau$, we can always choose a larger bandlimit $\Omega$ in order to make $\tau$ smaller, and therefore in principle evaluate $\mathcal{P}_{HOM}(\tau_H)$ for any value of $\tau_H$.

\begin{figure}
    \centering
    \includegraphics[width=0.45\textwidth]{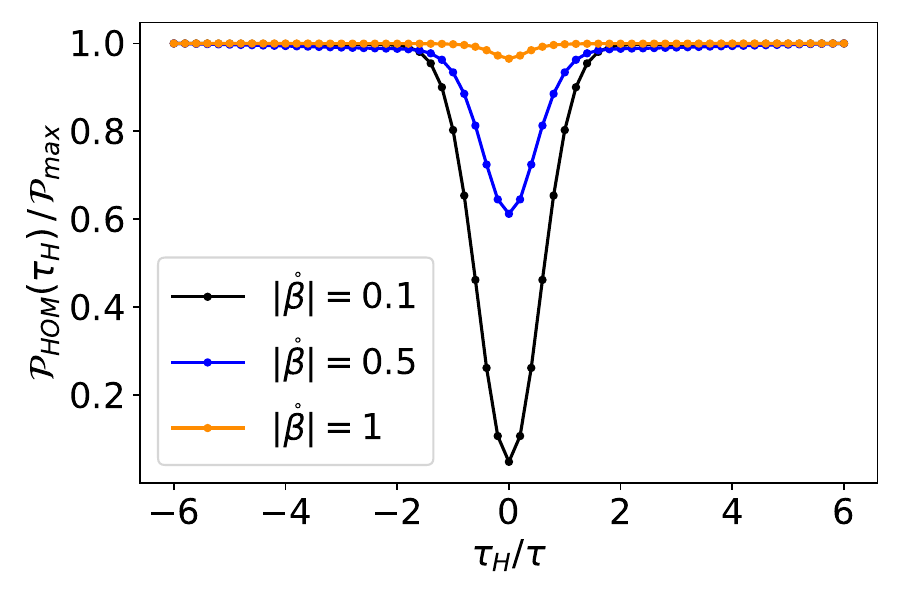}
    \caption{Normalized Hong-Ou-Mandel coincidence probability $\mathcal{P}_{HOM}(\tau_H)/\mathcal{P}_{max}$ vs. $\tau_H/\tau$ for the double Gaussian joint amplitude with $T_p/T_c=10$ at various values of $|\mathring{\beta}|$, in a time window containing the entire pulse. When the state is weakly squeezed the bottom of the dip approaches zero, and as $|\mathring{\beta}|$ is increased the dip becomes shallower. To obtain a higher resolution we increased the bandlimit $\Omega$ so that calculations were made with $\tau'=\tau/10$, but the $|\mathring{\beta}|$ and $\tau$ shown in the plot are those from the minimal bandlimit $\Omega=2\sqrt{\pi}/T_c$.}
    \label{HOM_Dip_Fig}
\end{figure}

For weakly squeezed light in a single pair window, the coincidence probability has the usual Hong-Ou-Mandel dip; since the detected photons are from the same pair, they are indistinguishable (given a symmetric joint amplitude) if there is no time delay \cite{DragoHOM}. As squeezing becomes stronger, some of the contribution to the coincidence probability is due to photons from different pairs, and these will not necessarily destructively interfere in one of the beam splitter outputs. We see in Fig.~\ref{HOM_Dip_Fig} that as $|\mathring{\beta}|$ increases (stronger squeezing), the minimum of the (normalized) dip becomes higher.

\begin{figure}
    \centering
    \includegraphics[width=0.45\textwidth]{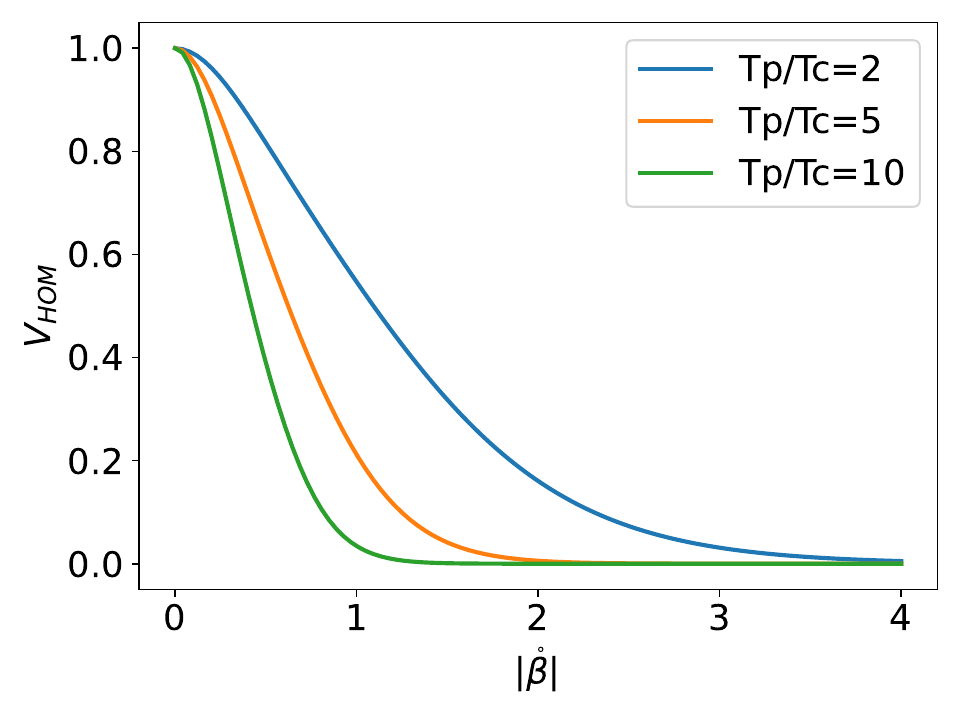}
    \caption{Hong-Ou-Mandel visibility $V_{HOM}$ vs. $|\mathring{\beta}|$ for the double Gaussian joint amplitude in a time window containing the entire pulse. Visibility decreases with $|\mathring{\beta}|$ since there is a greater contribution to the coincidence probability from separate pairs. The visibility degrades less for joint amplitudes corresponding to shorter pulse lengths.}
    \label{HOM_V_Fig}
\end{figure}

We can also examine the visibility of the HOM dip, defined as
\begin{equation}
    V_{HOM}=\frac{\mathcal{P}_{max}-\mathcal{P}_{min}}{\mathcal{P}_{max}},
\end{equation}
which we plot in Fig.~\ref{HOM_V_Fig} as a function of $|\mathring{\beta}|$. The visibility approaches unity when $|\mathring{\beta}|$ is small, and gets close to zero on the order of $|\mathring{\beta}|\approx1$. The presence of multiple pairs is known to lead to accidental coincidences that reduce the visibility \cite{Zhang_2010_HOM_experiment,HOM_2_Pair}, aligning with our analysis. Moreover, the visibility when accounting for multiple photon pairs is higher for more spectrally pure joint amplitudes, which was observed in similar HOM schemes that used two SPDC sources \cite{Jin_HOM,Thomas_HOM}. Given that a more spectrally pure squeezed vacuum has a shorter pulse length (compared to the coherence time) \cite{drago2023takingapartsqueezedlight}, signal and idler photons are more likely to arrive at the same time and be indistinguishable from one another.

The dip minimum can be obtained without shifting any indices since $\boldsymbol{\mathring{T}}^J=\boldsymbol{T}^J$ for $\tau_H=0$, and in the case where the joint amplitude is symmetric, it simplifies to
\begin{equation}
    \label{HOM sym min}
        \mathcal{P}^{sym}_{min}=(1-|\boldsymbol{W}^J|)^2.
\end{equation}
As shown in Appendix \ref{HOMDix}, for a finite (ie. \textit{not} CW) pulse contained entirely within $J$ we arrive at an expression for $\mathcal{P}_{max}\equiv\mathcal{P}_{HOM}(\tau\to\infty)$ without having to shift any matrix indices:
\begin{equation}
\label{HOM max}
    \begin{aligned}
        \mathcal{P}_{max}&=1+|\boldsymbol{W}|^2\bigg(1-2\big|\boldsymbol{I}-\frac{1}{4}\tanh^2\boldsymbol{Q}\big|^{-1}\bigg).
    \end{aligned}
\end{equation}
Ref. \cite{2-mode_HOM} finds expressions for the dip minimum and maximum of a two mode squeezed vacuum using a covariance matrix approach, and Eqs.~\eqref{HOM max} and \eqref{HOM sym min} reduce to the same result in the limit of two modes (when we set losses to zero). According to Ref. \cite{Jin_HOM}, the covariance matrix method can be extended to multimode squeezed light using the Schmidt decomposition, but this will not be applicable in the CW limit where our Whittaker-Shannon analysis can apply.

\section{Conclusion}

We have formulated the Whittaker-Shannon decomposition for nondegenerate squeezed light, found the disentangling formula of the multimode nondegenerate squeezing operator, and used the Whittaker-Shannon decomposition to analyze squeezed light in three detection schemes.  The quadrature variance reduction measured with homodyne detection justified our use of $\betac$ to quantify squeezing strength \cite{Strong_Optomechanical_Squeezing,Intense_Squeezing,10dB_strength}. In a polarization dependent coincidence detection scheme introduced by Takesue \cite{TAKESUE2010276}, the Whittaker-Shannon decomposition allowed us to find coincidence probabilities within a time window, even in the CW limit, and showed that entanglement between signal and idler photons has a greater effect on coincidence probabilities when squeezing is weak. Our expressions for the coincidence probabilities are more general than the previous analysis since they are compatible with any joint amplitude. The temporal properties of the Whittaker-Shannon modes also lead to expressions for the coincidence probability of multimode light of arbitrary squeezing strength in a Hong-Ou-Mandel scheme, and they can be applied in the CW limit where other results can not \cite{Takeoka_2015,Ferrari_HOM}. We find that the Hong-Ou-Mandel dip becomes shallower as squeezing strength increases.

Our examples show squeezed light exhibits its quantum nature in distinct ways. Effects due to the entanglement of photon pairs are strongest in the weakly squeezed regime, since any photons detected are likely to belong to the same pair. In contrast, quadrature noise can be squeezed the most when there are many photon pairs. Future work will investigate how the behaviour in these opposing regimes relates to different measures of nonclassicality. Negativity of the Wigner function does not detect any nonclassicality for squeezed light 
\cite{WignerFunction}. However, other measures indicate nonclassicality that increases with the strength of a single-mode squeezed state \cite{QCS, Nonclassical_Depth_1, Nonclassical_Distance}, and future work will investigate these measures for multimode squeezed light. The presence of multiple photon pairs reduces entanglement \cite{Brewster_polarization_entanglement_degradation,Adam_Multiphoton_Noise,Hošák2021_QKD,Kim_QKD} and degrades interference visibility in many schemes \cite{Zhang_2010_HOM_experiment,HOM_2_Pair,Zhong_Visibility,Nauth_Counting_statistics}. Thus, measures of entanglement could serve to describe regimes where quantum effects are greater for low squeezing strength.

A major advantage of the Whittaker-Shannon decomposition over the Schmidt decomposition is the ability to approximate the ket within the time window of interest. However, by reconstructing the joint amplitude within a finite time window, the Whittaker-Shannon decomposition could be used to define a temporally local Schmidt decomposition. This method may retain advantages of the Schmidt decomposition, including the Schmidt number as a measure of entanglement \cite{Measuring_Schmidt_Number}, while also having the Whittaker-Shannon formalism’s ability to pick out the state at relevant times.

\section{Acknowledgments}
J. E. Sipe acknowledges support from the Natural Sciences and Engineering Research Council of Canada (NSERC). J. Kranias acknowledges the support of an NSERC CGSM scholarship.

\FloatBarrier
\bibliography{bibliography}

\onecolumngrid

\begin{appendix}

\section{Disentangling of Nondegenerate Squeezing Operator}
\label{DisentanglingDix}
Here we outline the proof for the disentangling formula of the nondegenerate squeezing operator, which follows closely from the degenerate case in \cite{Disentangling}. Define the operators

\begin{equation} \begin{aligned}
    A(u)=\ol{b}u^\dagger a, \quad A^\dagger(v)=\ol{a}^\dagger v b^\dagger, \quad B(w,x)=\ol{a}^\dagger(wx^\dagger)a+\ol{b}(x^\dagger w)b^\dagger,
\end{aligned} \end{equation}
where $u,v,w,x$ are square matrices, but not in general symmetric as they are in the degenerate case. We also define $a=\begin{pmatrix}a_o&a_1&\ldots\end{pmatrix}^T$, $a^\dagger=\begin{pmatrix}a^\dagger_o&a^\dagger_1&\ldots\end{pmatrix}^T$ and let $\ol{a}$ ($\ol{a}^\dagger$) denote the transpose of $a$ ($a^\dagger$) ($b$ and $b^\dagger$ are defined analogously).  To prove that $\mathcal{L}\equiv\{A(u),A^\dagger(v),B(w,x) \;|\; u,v,w,x=(zz^\dagger)^nz,\; n=0,1,2,\ldots\}$ forms a Lie group, we must show that $[a,b]\in\mathcal{L}$ for all $a,b\in\mathcal{L}$. (The other Lie group axioms are satisfied automatically by the definition of the commutator.) Using Einstein notation to calculate the commutators:

\begin{equation} \begin{aligned}
    [A(u),A^\dagger(v)]&=[b_iu^\dagger_{ij}a_j, a^\dagger_kv_{kl}b^\dagger_l]\\
    &=u^\dagger_{ij}v_{kl}(b_i[a_j,a^\dagger_kb^\dagger_l]+[b_i,a^\dagger_kb^\dagger_l]a_j)\\
    &=u^\dagger_{ij}v_{kl}(b_ib^\dagger_l\delta_{jk}+a^\dagger_ka_j\delta_{il})\\
    &=a^\dagger_kv_{kl}u^\dagger_{lj}a_j+b_iu^\dagger_{ij}v_{jl}b^\dagger_l\\
    &=\ol{a}^\dagger(vu^\dagger)a+\ol{b}(u^\dagger v)b^\dagger\\
\end{aligned} \end{equation}
\begin{equation} \begin{aligned}
    [A(u),B(w,x)]=&[b_iu^\dagger_{ij}a_j,a^\dagger_kw_{kl}x^\dagger_{lm}a_m+b_kx^\dagger_{kl}w_{lm}b^\dagger_m]\\
    &=b_iu^\dagger_{ij}w_{kl}x^\dagger_{lm}a_m\delta_{jk}+b_ku^\dagger_{ij}x^\dagger_{kl}w_{lm}a_j\delta_{im}\\
    &=\ol{b}(u^\dagger wx^\dagger)a+\ol{b}(x^\dagger wu^\dagger)a
\end{aligned} \end{equation}
\begin{equation} \begin{aligned}
    [A^\dagger(v),B(w,x)]&=[a^\dagger_iv_{ij}b^\dagger_j,a^\dagger_kw_{kl}x^\dagger_{lm}a_m+b_kx^\dagger_{kl}w_{lm}b^\dagger_m]\\
    &=-a^\dagger_kw_{kl}x^\dagger_{lm}v_{mj}b^\dagger_j-a^\dagger_iv_{ij}x^\dagger_{jl}w_{lm}b^\dagger_m\\
    &=-\ol{a}^\dagger(wx^\dagger v)b^\dagger-\ol{a}^\dagger(vx^\dagger w)b^\dagger.
\end{aligned} \end{equation}
Therefore we have the commutation relations
\begin{equation} \begin{aligned}
    [A(u),A^\dagger(v)]&=B(v,u),\\
    [A(u),B(w,x)]&=A(xw^\dagger u)+A(uw^\dagger x), \\ 
    [A^\dagger(v),B(w,x)]&=-A^\dagger(wx^\dagger v)-A^\dagger(vx^\dagger w).
\end{aligned} \end{equation}
Now since $u,v,w,x=(zz^\dagger)^nz$, then $xw^\dagger u=(zz^\dagger)^nzz^\dagger(zz^\dagger)^m(zz^\dagger)^lz=(zz^\dagger)^{n+m+l+1}z$, so $A(xw^\dagger v)\in\mathcal{L}$, and similarly for $A(uw^\dagger x),A^\dagger(wx^\dagger v),A^\dagger(vx^\dagger w)$. Therefore $\mathcal{L}$ forms a Lie group. By Ado's theorem \cite{Disentangling}, we can construct a faithful matrix representation
\begin{equation} \begin{aligned}
    A(u)=\begin{pmatrix}
        0&0\\-u^\dagger&0
    \end{pmatrix},\quad A^\dagger(v)=\begin{pmatrix}
        0&v\\0&0
    \end{pmatrix},\quad B(w,x)=\begin{pmatrix}
        wx^\dagger&0\\0&-x^\dagger w
    \end{pmatrix},
\end{aligned} \end{equation}
which can be verified to obey the correct commutation relations. The operator we are interested in disentangling is
\begin{equation} \begin{aligned}
    S=e^{A^\dagger(z)-A(z)},
\end{aligned} \end{equation}
and mapping to the matrix form of the operators we have
\begin{equation} \begin{aligned}
    S&=e^{\begin{pmatrix}
        0&z\\z^\dagger&0
    \end{pmatrix}}\\
    &=\sum_{n=0}^\infty\frac{1}{n!}\begin{pmatrix}
        0&z\\z^\dagger&0
    \end{pmatrix}^n\\
    &=\sum_{n=0}^\infty\frac{1}{(2n)!}\begin{pmatrix}
        0&z\\z^\dagger&0
    \end{pmatrix}^{2n}+\sum_{n=0}^\infty\frac{1}{(2n+1)!}\begin{pmatrix}
        0&z\\z^\dagger&0
    \end{pmatrix}^{2n+1}\\
    &=\sum_{n=0}^\infty\frac{1}{(2n)!}\begin{pmatrix}
        (zz^\dagger)^n&0\\0&(z^\dagger z)^n\end{pmatrix}+\sum_{n=0}^\infty\frac{1}{(2n+1)!}\begin{pmatrix}
        0&(zz^\dagger)^nz\\(z^\dagger z)^nz^\dagger&0
    \end{pmatrix}.
\end{aligned} \end{equation}
The polar decomposition $z=UP=QU$ allows us to write
\begin{equation} \begin{aligned}
    (zz^\dagger)^n=(QUU^\dagger Q)^n=Q^{2n},\\
    (z^\dagger z)^n=(PU^\dagger UP)^n=P^{2n},\\
    (zz^\dagger)^nz=Q^{2n}QU=Q^{2n+1}U,\\
    (z^\dagger z)^nz^\dagger=z^\dagger(zz^\dagger)^n=U^\dagger Q^{2n+1},
\end{aligned} \end{equation}
and we can write $S$ as
\begin{equation} \begin{aligned}
    S&=\begin{pmatrix}
        \sum_{n=0}^\infty\frac{1}{(2n)!}Q^{2n}&0\\0&\sum_{n=0}^\infty\frac{1}{(2n)!}P^{2n}
    \end{pmatrix}+\begin{pmatrix}
        0&\sum_{n=0}^\infty\frac{1}{(2n+1)!}Q^{2n+1}U\\\sum_{n=0}^\infty\frac{1}{(2n+1)!}U^\dagger Q^{2n+1}&0
    \end{pmatrix}\\
    &=\begin{pmatrix}
        \cosh Q&(\sinh Q)U\\U^\dagger(\sinh Q)&\cosh P
    \end{pmatrix}.
\end{aligned} \end{equation}
We now seek a factorization of $S$ of the form
\begin{equation} \begin{aligned}
    \begin{pmatrix}
        \cosh Q&(\sinh Q)U\\U^\dagger(\sinh Q)&\cosh P
    \end{pmatrix}=\begin{pmatrix}
        I&\alpha\\0&I
    \end{pmatrix}\begin{pmatrix}
        \rho&0\\0&\gamma
    \end{pmatrix}\begin{pmatrix}
        I&0\\\delta&I
    \end{pmatrix}=\begin{pmatrix}
        \rho+\alpha\gamma\delta&\alpha\gamma\\\gamma\delta&\gamma
    \end{pmatrix},
\end{aligned} \end{equation}
where $\alpha,\beta,\gamma,\rho$ are square matrices. Therefore
\begin{equation} \begin{aligned}
    \gamma&=\cosh P
\end{aligned} \end{equation}
which implies
\begin{equation}\begin{aligned}
    \alpha&=(\tanh Q)U, & \delta=U^\dagger(\tanh Q)&=\alpha^\dagger,
\end{aligned}\end{equation}
and then
\begin{equation} \begin{aligned}
    \rho&=\cosh Q -(\tanh Q)U(\cosh P)U^\dagger(\tanh Q)=(\cosh Q)^{-1}.
\end{aligned} \end{equation}
If we write the matrix factors as matrix exponentials, then
\begin{equation} \begin{aligned}
    \begin{pmatrix}
        I&\alpha\\0&I
    \end{pmatrix}=\begin{pmatrix}
        I&0\\0&I
    \end{pmatrix}+\begin{pmatrix}
        0&\alpha\\0&0
    \end{pmatrix}=e^{\begin{pmatrix}
        0&\alpha\\0&0
    \end{pmatrix}},
\end{aligned} \end{equation}
and
\begin{equation} \begin{aligned}
    \begin{pmatrix}
        I&0\\\delta&I
    \end{pmatrix}=e^{\begin{pmatrix}
        0&0\\\delta&0
    \end{pmatrix}}=e^{\begin{pmatrix}
        0&0\\\alpha^\dagger&0
    \end{pmatrix}}.
\end{aligned} \end{equation}
For some $w,x^\dagger$, we should have
\begin{equation} \begin{aligned}
    \begin{pmatrix}
        \rho&0\\0&\gamma
    \end{pmatrix}=e^{\begin{pmatrix}
        wx^\dagger&0\\0&-x^\dagger w
    \end{pmatrix}}=\begin{pmatrix}
        e^{wx^\dagger}&0\\0&e^{-x^\dagger w}
    \end{pmatrix}.
\end{aligned} \end{equation}
Note that $\gamma=U^\dagger\rho^{-1}U$, therefore,
\begin{equation} \begin{aligned}
    \rho&=e^{wx^\dagger} & U^\dagger\rho^{-1}U&=e^{-x^\dagger w}\\
    \implies wx^\dagger&=\ln\rho & \implies x^\dagger w&=U^\dagger(\ln\rho) U,
\end{aligned} \end{equation}
which is solved by $w,x=(\ln\rho)^{\frac{1}{2}}U$. Now we can write $S$ and map the matrices back to operators:
\begin{equation} \begin{aligned}
    S&=e^{\begin{pmatrix}
        0&\alpha\\0&0
    \end{pmatrix}}e^{\begin{pmatrix}
        \ln\rho&0\\0&-U^\dagger(\ln\rho)U
    \end{pmatrix}}e^{\begin{pmatrix}
        0&0\\\alpha^\dagger&0
    \end{pmatrix}}\\
     &\to e^{A^\dagger(\alpha)}e^{B((\ln\rho)^{\frac{1}{2}}U,(\ln\rho)^{\frac{1}{2}}U)}e^{-A(\alpha)},
\end{aligned} \end{equation}
so the disentangled nondegenerate squeezing operator is
\begin{equation} \begin{aligned}
    S=e^{\ol{a}^\dagger z b^\dagger-\ol{b}z^\dagger a}=e^{\ol{a}^\dagger\alpha b^\dagger}e^{\ol{a}^\dagger(\ln\rho)a+\ol{b}(U^\dagger(\ln\rho)U)b^\dagger}e^{-\ol{b}\alpha^\dagger a},
\end{aligned} \end{equation}
or in terms of $Q$ and $U$,
\begin{equation} \begin{aligned}
    S=e^{\ol{a}^\dagger(\tanh Q)Ub^\dagger}e^{-\ol{a}^\dagger\ln(\cosh Q)a-\ol{b}U^\dagger(\ln(\cosh Q))Ub^\dagger}e^{-\ol{b}U^\dagger(\tanh Q) a}.
\end{aligned} \end{equation}
Letting $M=U^\dagger(\ln(\cosh Q))U$, we normally order the middle term:
\begin{equation} \begin{aligned}
    \ol{b}Mb^\dagger=b_iM_{ij}b^\dagger_j=M_{ij}(b^\dagger_jb_i+\delta_{ij})=\ol{b}^\dagger M^Tb+\tr(M).
\end{aligned} \end{equation}
Now noting that $M^T=\ln(\cosh P^T)$ and $e^{\tr(\ln M)}=|M|$ we find
\begin{equation} \begin{aligned}
    S&=|\sech P|e^{\ol{a}^\dagger(\tanh Q)Ub^\dagger}e^{-\ol{a}^\dagger\ln(\cosh Q)a-\ol{b}^\dagger\ln(\cosh P^T)b}e^{-\ol{b}U^\dagger(\tanh Q) a}.
\end{aligned} \end{equation}
When acting on the vacuum, the last two terms will give $e^{0}=1$, so
\begin{equation} \begin{aligned}
    S\ket{\text{vac}}=|\sech P|e^{\ol{a}^\dagger(\tanh Q)Ub^\dagger}\ket{\text{vac}}.
\end{aligned} \end{equation}
And since $|\cosh P|=|U^\dagger(\cosh Q)U|=|\cosh Q|$, we can write the state only in terms of $Q$ and $U$:
\begin{equation} \begin{aligned}
    S\ket{\text{vac}}&=|\sech Q|e^{\ol{a}^\dagger(\tanh Q)Ub^\dagger}\ket{\text{vac}}.
\end{aligned} \end{equation}
When applying this disentangling formula to Eq.~\eqref{Whittaker-Shannon Ket} we set $z=\boldsymbol{\beta}$ and have the same polar decomposition.

\section{Calculation of Moments}
\label{Moments Dix}

Here we explain how to obtain the $N$ and $M$ moments. The moments of degenerate squeezed light are found with the methods in \cite{drago2023takingapartsqueezedlight}. Here we show similar derivations for nondegenerate squeezed light. 

First we find the transformation induced by the nondegenerate squeezing operator $S=e^{\beta_{nm}A_n^\dagger B_m^\dagger-h.c.}$ on the Whittaker-Shannon mode operators. Let $X=-\beta_{nm}A_n^\dagger B_m^\dagger+h.c.$, so that $S=e^{-X}$, and the operator expansion theorem tells us that \cite{Gerry_Knight_2004}
\begin{equation} \begin{aligned}
    S^\dagger A_r S=A_r+[X,A_r]+\frac{1}{2!}[X,[X,A_r]]+\hdots.
\end{aligned} \end{equation}
Now compute the successive commutators:
\begin{equation} \begin{aligned}
    [X,A_r]&=-\beta_{nm}B_m^\dagger[A_n^\dagger,A_r]=\beta_{nm}B_m^\dagger\delta_{nr}=\beta_{rs}B^\dagger_s,\\
    [X,[X,A_r]]&=[X,\beta_{rs}B_s^\dagger]=\beta^*_{nm}\beta_{rs}A_n[B_m,B_s^\dagger]=\beta^*_{nm}\beta_{rs}A_n\delta_{ms}=\beta_{ra}\beta^\dagger_{as}A_s,\\
    [X,[X,[X,A_r]]]&=[X,\beta_{ra}\beta^*_{sa}A_s]=\beta_{ra}\beta^*_{sa}\beta_{nm}B_m^\dagger\delta_{ns}=\beta_{ra}\beta^\dagger_{ab}\beta_{bs}B_s^\dagger,\\
    [X,[X,[X,[X,A_r]]]]&=[X,\beta_{ra}\beta^*_{ba}\beta_{bs}B_s^\dagger]=\beta_{ra}\beta^*_{ba}\beta_{bs}\beta^*_{nm}A_n\delta_{ms}=\beta_{ra}\beta^\dagger_{ab}\beta_{bc}\beta^\dagger_{cs}A_s.
\end{aligned} \end{equation}
Each iteration switches between $A_s$ (even) and $B_s^\dagger$ (odd), and adds a factor $\beta^\dagger_{is}$ ($\beta_{is}$) for even (odd) terms, where $i$ is the last index of the previous factor. Therefore, we can write
\begin{equation} \begin{aligned}
    S^\dagger A_r S&=\mu^A_{rs}A_s+\nu^A_{rs}B_s^\dagger\\
    \mu^A_{rs}&=\delta_{rs}+\frac{1}{2!}\beta_{ra}\beta^\dagger_{as}+\frac{1}{4!}\beta_{ra}\beta^\dagger_{ab}\beta_{bc}\beta^\dagger_{cs}+\hdots\\
    \nu^A_{rs}&=\beta_{rs}+\frac{1}{3!}\beta_{ra}\beta^\dagger_{ab}\beta_{bs}+\frac{1}{5!}\beta_{ra}\beta^\dagger_{ab}\beta_{bc}\beta^\dagger_{cd}\beta_{ds}+\hdots.
\end{aligned} \end{equation}
The above can be expressed in matrix form with the polar decomposition of $\boldsymbol{\beta}$:
\begin{equation} \begin{aligned}
    \boldsymbol{\mu}^A&=I+\frac{1}{2!}\boldsymbol{\beta}\boldsymbol{\beta}^\dagger+\frac{1}{4!}\boldsymbol{\beta}\boldsymbol{\beta}^\dagger\boldsymbol{\beta}\boldsymbol{\beta}^\dagger+\hdots=I+\frac{1}{2!}\boldsymbol{U}\boldsymbol{P}^2\boldsymbol{U}^\dagger+\hdots=\cosh\boldsymbol{Q}\\
    \boldsymbol{\nu}^A&=\boldsymbol{\beta}+\frac{1}{3!}\boldsymbol{\beta}\boldsymbol{\beta}^\dagger\boldsymbol{\beta}+\frac{1}{5!}\boldsymbol{\beta}\boldsymbol{\beta}^\dagger\boldsymbol{\beta}\boldsymbol{\beta}^\dagger\boldsymbol{\beta}+\hdots=\boldsymbol{U}\boldsymbol{P}+\frac{1}{3!}\boldsymbol{U}\boldsymbol{P}^3+\hdots=(\sinh\boldsymbol{Q})\boldsymbol{U}.
\end{aligned} \end{equation}
Note that $\boldsymbol{\mu}^A$ is Hermitian, but $\boldsymbol{\nu}^A$ is not symmetric since $\boldsymbol{\beta}$ is not symmetric.
To find $S^\dagger B_rS$ we can re-index the sum in the exponent of $S$ to find
\begin{equation} \begin{aligned}
    S^\dagger B_rS&=e^{-\beta_{nm}A^\dagger_nB^\dagger_m-h.c}B_re^{\beta_{nm}A^\dagger_nB^\dagger_m+h.c}\\
    &=e^{-\beta_{nm}B^\dagger_mA^\dagger_n-h.c}B_re^{\beta_{nm}B^\dagger_mA^\dagger_n+h.c}\\
    &=e^{-\beta^T_{nm}B^\dagger_nA^\dagger_m-h.c}B_re^{\beta^T_{nm}B^\dagger_nA^\dagger_m+h.c},
\end{aligned} \end{equation}
meaning that the transformation on $B_r$ is the same as on $A_r$ but with $\boldsymbol{\beta}\to\boldsymbol{\beta}^T$, which is equivalent to taking
\begin{equation} \begin{aligned}
    \boldsymbol{U}\to \boldsymbol{U}^T,\quad \boldsymbol{Q}\to \boldsymbol{P}^T,\quad \boldsymbol{P}\to \boldsymbol{Q}^T.
\end{aligned} \end{equation}
Therefore,
\begin{equation} \begin{aligned}
    S^\dagger B_rS&=\mu^B_{rs}B_s+\nu^B_{rs}A_s^\dagger,\\
    \boldsymbol{\mu}^B&=\cosh \boldsymbol{P}^T=\cosh (\boldsymbol{U}^\dagger \boldsymbol{Q}\boldsymbol{U})^T=(\boldsymbol{U}^\dagger \boldsymbol{\mu}^A\boldsymbol{U})^T,\\
    \boldsymbol{\nu}^B&=(\sinh \boldsymbol{P}^T)\boldsymbol{U}^T=\boldsymbol{U}^T(\sinh \boldsymbol{Q}^T)=(\boldsymbol{\nu}^A)^T.
\end{aligned} \end{equation}
Using these transformations, we calculate
\begin{equation}
    \begin{aligned}
        N^a_{nm}&=\bra{\psi}A^\dagger_nA_m\ket{\psi}\\
        &=\bra{\vac}S^\dagger A^\dagger_nA_mS\ket{\psi}\\
        &=\bra{\text{vac}}((\mu^A)^*_{na}A^\dagger_a+(\nu^A)^*_{na}B_a(\mu^A_{mb}A_b+\nu^A_{mb}B^\dagger_b)\ket{\text{vac}}\\
        &=(\nu^A)^*_{na}\nu^A_{mb}\delta_{ab}\\
        &=(\nu^A)^*_{na}\nu^A_{ma}\\
        &=((\boldsymbol{\nu}^A)^*(\boldsymbol{\nu}^A)^T)_{nm}\\
        &=((\sinh\boldsymbol{Q}^*)\boldsymbol{U}^*\boldsymbol{U}^T(\sinh\boldsymbol{Q}^T))_{nm}=((\sinh\boldsymbol{Q}^T)(\sinh\boldsymbol{Q}^T))_{nm}\\
        &=(\sinh^2\boldsymbol{Q}^T)_{nm},
    \end{aligned}
\end{equation}
and similarly for the other moments. The moments of the CT modes are found in terms of those of the Whittaker-Shannon supermodes using Eq.~\eqref{invertedWSmodes}, for example,
\begin{equation}
    \begin{aligned}
        N^a(t_1,t_2)&=\bra{\psi}\ol{a}^\dagger(t_1)\ol{a}(t_2)\ket{\psi}\\
        &=\ol{\chi}^*_n(t_1)\bra{\psi}A^\dagger_nA_m\ket{\psi}\ol{\chi}_m(t_2)\\
        &=\ol{\chi}^*_n(t_1)N^a_{nm}(t_2)\ol{\chi}_m(t_2).
    \end{aligned}
\end{equation}

\section{Homodyne Spectral Analysis}
\label{CW Dix}
This appendix shows the derivation of the quadrature variance of homodyne detection in the CW limit. Evaluating Eq.~\eqref{Current Covariance} gives
\begin{equation}
    \begin{aligned}
    \langle i(t)i(t+\tilde{\tau})\rangle&=|\eta|^2\big(\delta(\tilde{\tau})+N^d(t,t+\tilde{\tau})+N^d(t+\tilde{\tau},t)\\
    &\quad+e^{2i\theta}M^d(t,t+\tilde{\tau})+e^{-2i\theta}\big(M^d(t+\tilde{\tau},t)\big)^*\big)+\delta(\tilde{\tau})G^{(n)}(t,t+\tilde{\tau}),
    \end{aligned}
\end{equation}
but if the local oscillator is much stronger than the signal, the last term will be insignificant since it is not proportional to $|\eta|^2$ like the others. Integrating over $t$ and $\tilde{\tau}$ results in
\begin{equation}
    \begin{aligned}
        \int_J dt\int d\tilde{\tau}N^d(t,t+\tilde{\tau})e^{-i\omega\tilde{\tau}}&=N^d_{nm}\int_J dt\int d\tilde{\tau}\ol{\chi}_n(t)\ol{\chi}_m(t+\tilde{\tau})e^{-i\omega\tilde{\tau}}\\
        &=\frac{1}{\sqrt{2\pi}}N^d_{nm}\int_J dt\ol{\chi}_n(t)\int d\tilde{\tau}\int d\omega'\chi_m(\omega')e^{-i\omega'(t+\tilde{\tau})}e^{-i\omega\tilde{\tau}}\\
        &=\sqrt{2\pi}N^d_{nm}\int_J dt\ol{\chi}_n(t)\int d\omega'\chi_m(\omega')e^{-i\omega't}\delta(\omega+\omega')\\
        &=\sqrt{2\pi}N^d_{nm}\int_J dt\ol{\chi}_n(t)e^{i\omega t}\chi_m(-\omega)\\
        &\approx2\pi N^d_{nm}\chi_n(\omega)\chi_m(-\omega)\\
        &=\tau N^d_{nm}e^{i\omega(n-m)\tau}
    \end{aligned}
\end{equation}
by assuming in the second last line that $\int_Jdt\ol{\chi}_n(t)e^{i\omega t}=\chi_n(\omega)$. This is a good approximation when $T\gg\tau$ and $n\tau$ is sufficiently far from the edges of the time window. Since we are summing over all $n\tau$ within the time window, some edge modes where this expression is not valid will be included, but we can neglect their effects if $T$ is large enough. When $\betac$ is larger it may be necessary to increase $T$ when calculating the variance. The integrals work out similarly for the other terms, and performing the sums over $n$ and $m$ results in Eq.~\eqref{V CW}.

\section{Fredholm Equation to Matrix Equation}
Here we show how approximate completeness of the Whittaker-Shannon modes is used to convert the Fredholm integral equation into a matrix eigenvalue equation. Since $\ol{\chi}_n(t)=\ol{\chi}_n^*(t)$, inserting Eq.~\eqref{covariance functions} into Eq.~\eqref{Fredholm Eqn} gives
\begin{equation}
   \begin{aligned}
       \int_Jdt'\ol{\chi}_j(t)\ol{\chi}_k(t')\begin{pmatrix}
           (N^d_{jk})_R+(M^d_{jk})_R & (N^d_{jk})_I+(M^d_{jk})_I\\
           (M^d_{jk})_I-(N^d_{jk})_I & (N^d_{jk})_R-(M^d_{jk})_R
       \end{pmatrix}\ol{\chi}_l(t')\boldsymbol{\phi}_{nl}&=\lambda_n\ol{\chi}_l(t)
    \boldsymbol{\phi}_{nl},\\
    \ol{\chi}_j(t)\begin{pmatrix}
           (N^d_{jk})_R+(M^d_{jk})_R & (N^d_{jk})_I+(M^d_{jk})_I\\
           (M^d_{jk})_I-(N^d_{jk})_I & (N^d_{jk})_R-(M^d_{jk})_R
       \end{pmatrix}\boldsymbol{\phi}_{nk}&=\lambda_n\ol{\chi}_l(t)
    \boldsymbol{\phi}_{nl},\\
    \begin{pmatrix}
           (N^d_{jk})_R+(M^d_{jk})_R & (N^d_{jk})_I+(M^d_{jk})_I\\
           (M^d_{jk})_I-(N^d_{jk})_I & (N^d_{jk})_R-(M^d_{jk})_R
       \end{pmatrix}\boldsymbol{\phi}_{nk}&=\lambda_n
    \boldsymbol{\phi}_{nj},
    \end{aligned}
\end{equation}
where in the second line we used the argument that $\int_Jdt'\ol{\chi}_k(t')\ol{\chi}_l(t')=\delta_{kl}$ when $T\gg\tau$. This is equivalent to Eq.~\eqref{Matrix Eig Eqn}.

\section{Time Window Projectors}
\label{Projector Dix}
Here we prove that the projector we defined in Eq.~\eqref{projector} is indeed a projection operator. We make use of the identity
\begin{equation}
\label{perm_id}
    \begin{aligned}
        \bra{\vac}a_H(t_1)\hdots a_H(t_n)a_H^\dagger(t'_1)\hdots a_H^\dagger(t'_{n'})\ket{\vac}=\delta_{nn'}\sum_{\sigma(\{t'_n\})}\prod_{i=1}^n\delta(t_i-\sigma_i),
    \end{aligned}
\end{equation}
where $\sigma(\{t'_n\})$ is a permutation of the set $\{t'_n\}=\{t'_1,\hdots,t'_n\}$, $\sigma_i$ is the $i$th element of $\sigma(\{t'_n\})$, and the sum is over all permutations.

We can show that $P^{a_H}_{J,s}P^{a_H}_{J,s'}=\delta_{ss'}P^{a_H}_{J,s}$ as follows:
\begin{equation}
    \begin{aligned}
        P^{a_H}_{J,s}P^{a_H}_{J,s'}&=\frac{1}{s!s'!}\int_J dt_1\hdots dt_sdt'_1\hdots dt'_{s'}a_H^\dagger(t_1)\hdots a_H^\dagger(t_s)V_{a_H}^ J\\
        &\quad\times a_H(t_1)\hdots a_H(t_s)a_H^\dagger(t'_1)\hdots a_H^\dagger(t'_{s'}) V_{a_H}^ Ja_H(t'_1)\hdots a_H(t'_{s'})\\
        &=\frac{1}{s!s'!}\int_J dt_1\hdots dt_sdt'_1\hdots dt'_{s}a_H^\dagger(t_1)\hdots a_H^\dagger(t_s)V_{a_H}^ J\bigg(\delta_{ss'}\sum_{\sigma(\{t'_s\})}\prod_{i=1}^s\delta\big(t_i-\sigma_i\big)\bigg)a_H(t'_1)\hdots a_H(t'_{s'})\\
        &=\frac{\delta_{ss'}}{(s!)}\int_J dt_1\hdots dt_sdt'_1\hdots dt'_{s}a_H^\dagger(t_1)\hdots a_H^\dagger(t_s)V_{a_H}^ J\bigg(\prod_{i=1}^s\delta(t_i-t'_i)\bigg)a_H(t'_1)\hdots a_H(t'_{s})
        \\
        &=\frac{\delta_{ss'}}{(s!)}\int_J dt_1\hdots dt_sa_H^\dagger(t_1)\hdots a_H^\dagger(t_s) V_{a_H}^ Ja_H(t_1)\hdots a_H(t_{s})\\
        &=\delta_{ss'}P^{a_H}_{J,s},
\end{aligned}\end{equation}
since there are $s!$ permutations of the set $\{t'_n\}$ and we can exchange the order of the annihilation operators $a_H(t_i)$.

Now we must show that $\sum_sP^{a_H}_{J,s}$ is equal to the identity operator. We can write an arbitrary state in the $a_H$ Hilbert space as
\begin{equation}
    \ket{\varphi}=\sum_{n=0}^\infty p_n\ket{\varphi}_n,
\end{equation}
where $\sum_{n=0}^\infty|p_n|^2=1$. We set $\ket{\varphi}_0=\ket{\vac}_{a_H}$, and for $n>0$ $\ket{\varphi}_n$ is the $n$ photon ket
\begin{equation}
    \ket{\varphi}_n=\int dt_1\hdots dt_n\phi^{(n)}(t_1,\hdots,t_n)a_H^\dagger(t_1)\hdots a_H^\dagger(t_n )\ket{\vac}_{a_H}
\end{equation}
for some complex-valued function $\phi^{(n)}$ normalized by $\int dt_1\hdots dt_n|\phi^{(n)}(t_1,\hdots,t_n)|^2=1$. The multiple integral over all time can be decomposed as combinations of integrals over the window $J$ and the remaining times $\notJ$:
\begin{equation}
    \begin{aligned}
        \int dt_1\hdots dt_n=\left(\int_J dt_1 + \int_\notJ dt_1\right)\hdots\left(\int_J dt_n + \int_\notJ dt_n\right)=\sum_{k=0}^n\frac{1}{k!(n-k)!}\sum_{\sigma(\{t_n\})}\int_Jd\sigma_1\hdots d\sigma_k\int_\notJ d\sigma_{k+1}d\sigma_n,
    \end{aligned}
\end{equation}
so $\ket{\varphi}_n$ can expressed in a form where every creation operator is either in $J$ or $\notJ$:
\begin{equation}
    \begin{aligned}
        \ket{\varphi}_n&=\sum_{k=0}^n\frac{1}{k!(n-k)!}\sum_{\sigma(\{t_n\})}\int_Jd\sigma_1\hdots d\sigma_k\int_\notJ d\sigma_{k+1}d\sigma_n\phi^{(n)}(t_1,\hdots,t_n)a_H^\dagger(t_1)\hdots a_H^\dagger(t_n )\ket{\vac}_{a_H}\\
        \ket{\varphi}_n&=\sum_{k=0}^n\frac{1}{k!(n-k)!}\sum_{\sigma(\{t_n\})}\int_Jd\sigma_1\hdots d\sigma_k\int_\notJ d\sigma_{k+1}d\sigma_n\phi^{(n)}(t_1,\hdots,t_n)\\
        &\quad\quad\left(a_H^\dagger(\sigma_1)\hdots a_H^\dagger(\sigma_k )\ket{\vac}^J_{a_H}\right)\otimes\left(a_H^\dagger(\sigma_{k+1})\hdots a_H^\dagger(\sigma_n)\ket{\vac}^\notJ_{a_H}\right).
    \end{aligned}
\end{equation}
We could exchange the order of the creation operators to write them in terms of $\sigma$, but the arguments of $\phi^{(n)}$ stay as $\{t_n\}$ since their order matters.
\begin{equation}
    \begin{aligned}
        P^{a_H}_{J,s}\ket{\varphi}_n&=\frac{1}{s!}\sum_{k=0}^n\frac{1}{k!(n-k)!}\sum_{\sigma'\in\{t'_n\}}\int_Jdt_1\hdots dt_s\int_Jd\sigma'_1\hdots d\sigma'_k\int_\notJ d\sigma'_{k+1}d\sigma'_n \phi^{(n)}(t'_1,\hdots,t'_n)\\
        &\quad \quad \left(a_H^\dagger(t_1)\hdots a_H^\dagger(t_s)\ket{\vac}^J_{a_H}\bra{\vac}^J_{a_H} a_H(t_1)\hdots a_H(t_s)a_H^\dagger(\sigma'_1)\hdots a_H^\dagger(\sigma'_k)\ket{\vac}^J_{a_H}\right)\otimes\left(a_H^\dagger(\sigma'_{k+1})\hdots a_H^\dagger(\sigma'_n)\ket{\vac}^\notJ_{a_H}\right)\\  
        &=\frac{1}{s!}\sum_{k=0}^n\frac{1}{k!(n-k)!}\sum_{\sigma'\in\{t'_n\}}\int_Jdt_1\hdots dt_s\int_Jd\sigma'_1\hdots d\sigma'_k\int_\notJ d\sigma'_{k+1}d\sigma'_n \phi^{(n)}(t'_1,\hdots,t'_n)\\
        &\quad \quad \delta_{sk}\sum_{\sigma(\{t_s\})}\prod_{i=1}^s\delta(\sigma_i-\sigma'_i)\left(a_H^\dagger(t_1)\hdots a_H^\dagger(t_s)\ket{\vac}^J_{a_H}\right)\otimes\left(a_H^\dagger(\sigma'_{k+1})\hdots a_H^\dagger(\sigma'_n)\ket{\vac}^\notJ_{a_H}\right)\\
    \end{aligned}
\end{equation}
If $s>n$, the above will be zero since $s$ and $k$ can never be equal, if $s\leq n$ we exchange the order of $\{t_n\}$ so that $t_i=\sigma_i$ and find
\begin{equation}
    \begin{aligned}  
        P^{a_H}_{J,s}\ket{\varphi}_n&=\frac{1}{(s!)^2(n-s)!}\sum_{\sigma(\{t_s\})}\sum_{\sigma'\in\{t'_n\}}\int_Jd\sigma_1\hdots d\sigma_s\int_Jd\sigma'_1\hdots d\sigma'_s\int_\notJ d\sigma'_{s+1}d\sigma'_n \phi^{(n)}(t'_1,\hdots,t'_n)\\
        &\quad \quad \prod_{i=1}^s\delta(\sigma_i-\sigma'_i)\left(a_H^\dagger(\sigma_1)\hdots a_H^\dagger(\sigma_s)\ket{\vac}^J_{a_H}\right)\otimes\left(a_H^\dagger(\sigma'_{s+1})\hdots a_H^\dagger(\sigma'_n)\ket{\vac}^\notJ_{a_H}\right)\\
        &=\frac{1}{(s!)^2(n-s)!}\sum_{\sigma(\{t_s\})}\sum_{\sigma'\in\{t'_n\}}\int_Jd\sigma'_1\hdots d\sigma'_s\int_\notJ d\sigma'_{s+1}d\sigma'_n \phi^{(n)}(t'_1,\hdots,t'_n)\\
        &\quad \quad \left(a_H^\dagger(\sigma'_1)\hdots a_H^\dagger(\sigma'_s)\ket{\vac}^J_{a_H}\right)\otimes\left(a_H^\dagger(\sigma'_{s+1})\hdots a_H^\dagger(\sigma'_n)\ket{\vac}^\notJ_{a_H}\right)\\
        &=\frac{1}{s!(n-s)!}\sum_{\sigma'\in\{t'_n\}}\int_Jd\sigma'_1\hdots d\sigma'_s\int_\notJ d\sigma'_{s+1}d\sigma'_n \phi^{(n)}(t'_1,\hdots,t'_n)\\
        &\quad \quad \left(a_H^\dagger(\sigma'_1)\hdots a_H^\dagger(\sigma'_s)\ket{\vac}^J_{a_H}\right)\otimes\left(a_H^\dagger(\sigma'_{s+1})\hdots a_H^\dagger(\sigma'_n)\ket{\vac}^\notJ_{a_H}\right)\\.
    \end{aligned}
\end{equation}
Acting with $\sum_sP^{a_H}_{J,s}$ on the arbitrary state results in
\begin{equation}
    \begin{aligned}
        \sum_{s=0}^\infty P^{a_H}_{J,s}\ket{\varphi}&=\sum_{s=0}^\infty\sum_{n=0}^\infty p_n P^{a_H}_{J,s}\ket{\varphi}_n\\
        &=\sum_{n=0}^\infty p_n \sum_{s=0}^n \frac{1}{s!(n-s)!}\sum_{\sigma'\in\{t'_n\}}\int_Jd\sigma'_1\hdots d\sigma'_s\int_\notJ d\sigma'_{s+1}d\sigma'_n \phi^{(n)}(t'_1,\hdots,t'_n)\\
        &\quad \quad \left(a_H^\dagger(\sigma'_1)\hdots a_H^\dagger(\sigma'_s)\ket{\vac}^J_{a_H}\right)\otimes\left(a_H^\dagger(\sigma'_{s+1})\hdots a_H^\dagger(\sigma'_n)\ket{\vac}^\notJ_{a_H}\right)\\
        &=\sum_{n=0}^\infty p_n\ket{\varphi}_n=\ket{\varphi},
    \end{aligned}
\end{equation}
therefore, $\sum_sP^{a_H}_{J,s}$ is equal to the identity.

\section{Coincidence Probability Calculations}
\label{LargeTDix}
Here we show the calculation of coincidence probabilities when we have a large time window with $T\gg\tau$. Using Eq.~\eqref{invertedWSmodes}, we can rewrite the projector $P^{a_H}_{J,s}$ as
\begin{equation}
    \begin{aligned}
        P^{a_H}_{J,s}&=\frac{1}{s!}\sum_{n_1,\hdots,n_s}\sum_{m_1,\hdots,m_s} A^{H\dagger}_{n_1}\hdots A^{H\dagger}_{n_s} V_{a_H}^ JA^H_{m_1}\hdots A^H_{m_s}\\
        &\quad\times \int_J dt_1\hdots dt_s\ol{\chi}^*_{n_1}(t_1)\hdots\ol{\chi}^*_{n_s}(t_s)\ol{\chi}_{m_1}(t_1)\hdots\ol{\chi}_{m_s}(t_s)\\
        &=\frac{1}{s!}\sum_{n_1,\hdots,n_s}\sum_{m_1,\hdots,m_s} A^{H\dagger}_{n_1}\hdots A^{H\dagger}_{n_s} V_{a_H}^ JA^H_{m_1}\hdots A^H_{m_s}\\
        &\quad\times \bigg(\int_Jdt_1\ol{\chi}^*_{n_1}(t_1)\ol{\chi}_{m_1}(t_1)\bigg)\hdots\bigg(\int_Jdt_s\ol{\chi}^*_{n_s}(t_s)\ol{\chi}_{m_s}(t_s)\bigg).
    \end{aligned}
\end{equation}
Since $T\gg\tau$, the Whittaker-Shannon modes will be approximately orthonormal inside the time window, and we neglect the modes outside the window. Each integral over $t_i$ results in $\delta_{n_im_i}$, but only for the indices $n_i$ where $n_i\tau$ is inside the time window. Then we approximate the set of modes corresponding to $\ket{\vac}_{a_H}^ J$ as the set of Whittaker-Shannon modes for the relevant indices, and write the projector as

\begin{equation}
    P^{A^H}_{J,s}=\frac{1}{s!}\sum^J_{n_1,\hdots,n_s}A^{H\dagger}_{n_1}\hdots A^{H\dagger}_{n_s} V_{A^H}^JA^H_{n_1}\hdots A^H_{n_s},
\end{equation}
where the primed sum indicates we are only taking the indices which correspond to times within the Window, and $\ket{\vac}^J_{A^H}$ corresponds to the Whittaker-Shannon modes labeled by those indices. By using the discrete analogue of Eq.~\eqref{perm_id}
\begin{equation}
    \begin{aligned}
        \bra{\vac}A^H_{k_1}\hdots A^H_{k_n}A^{H\dagger}_{k'_1}\hdots A^{H\dagger}_{k'_{n\prime}}\ket{\vac}=\delta_{nn'}\sum_{\sigma(\{k'_n\})}\prod_{i=1}^n\delta_{k_i\sigma_i},
    \end{aligned}
\end{equation}
we can prove with similar steps as the previous section that $P^{A^H}_{J,s}P^{A^H}_{J,s'}=\delta_{ss'}P^{A^H}_{J,s}$ and $\sum_{s=0}^\infty P^{A^H}_{J,s}=\mathbb{I}_{A^H}^J$. Now to find the coincidence probabilities we sum over every combination of projectors where there is at least one signal and one idler photon, weighted by the detection probability of each number:
\begin{equation}
    \begin{aligned}
        \mathcal{P}_{HH}&=\sum_{s_a,s_b=1}^\infty D_{s_a}D_{s_b}\bra{\psi_H}P^{a_H}_{J,s_a}P^{b_H}_{J,s_b}\ket{\psi_H}\approx\sum_{s_a,s_b=1}^\infty D_{s_a}D_{s_b}\bra{\psi_H}P^{A^H}_{J,s_a}P^{B^H}_{J,s_b}\ket{\psi_H},\\
        \mathcal{P}_{HV}&=\sum_{s_a,s_b=1}^\infty D_{s_a}D_{s_b}\bra{\psi_H}P^{a_H}_{J,s_a}\ket{\psi_H}\bra{\psi_V}P^{b_V}_{J,s_b}\ket{\psi_V}\approx\sum_{s_a,s_b=1}^\infty D_{s_a}D_{s_b}\bra{\psi_H}P^{A^H}_{J,s_a}\ket{\psi_H}\bra{\psi_V}P^{B^V}_{J,s_b}\ket{\psi_V}
    \end{aligned}
\end{equation} 
However, we can see from the disentangled form of $\ket{\psi_H}$ that all terms have an equal number of signal and idler photons, therefore, $\bra{\psi_H}P^{a_H}_{J,s_a}P^{b_H}_{J,s_b}\ket{\psi_H}$ is only nonzero when $s_a=s_b$ and we have
\begin{equation}\begin{aligned}
    \mathcal{P}_{HH}&=\sum_{s=1}^\infty D_s^2\bra{\psi_H}P^{A^H}_{J,s}P^{B^H}_{J,s}\ket{\psi_H}, & \mathcal{P}_{HV}&=\sum_{s_a,s_b=1}^\infty D_{s_a}D_{s_b}\bra{\psi_H}P^{A^H}_{J,s_a}\ket{\psi_H}\bra{\psi_V}P^{B^V}_{J,s_b}\ket{\psi_V}.
\end{aligned}\end{equation}

The photon number probabilities can be calculated using the disentangled form of the ket. First let us consider $\bra{\psi_H}P^{A^H}_{J,s}P^{B^H}_{J,s}\ket{\psi_H}$. The only term in the expansion of $\ket{\psi_H}=| \boldsymbol{W}^J|e^{T_{jk}{A^H_j}^\dagger{B^H_k}^\dagger}\ket{\vac}$ that contributes is the one with s signal and idler photons, $\frac{1}{s!}(T_{jk}A^{H\dagger}_jB^{H\dagger}_k)^s\ket{\vac}_H^J$, where $\ket{\vac}_H^J\equiv\ket{\vac}_{A^H}^J\otimes\ket{\vac}_{B^H}^J$. Expanding, we find
\begin{equation}
    \begin{aligned}
        &\bra{\psi_H}P^{A^H}_{J,s}P^{B^H}_{J,s}\ket{\psi_H}\\
        &=\frac{| \boldsymbol{W}^J|^2}{(s!)^2}\sum^J_{n_1,\hdots, n_s}\sum^J_{m_1,\hdots,m_s}\bigg|\frac{1}{s!}\sum^J_{j_1,\hdots,j_s}\sum^J_{k_1,\hdots,k_s}T_{j_1k_1}\hdots T_{j_sk_s}\bra{\vac}_H^J A^{H\dagger}_{n_1}  B^{H\dagger}_{m_1}\hdots  A^{H\dagger}_{n_s}  B^{H\dagger}_{m_s} A^H_{j_1}B^H_{k_1}\hdots A^H_{j_s}B^H_{k_s}\ket{\vac}_H^J\bigg|^2.
    \end{aligned}
\end{equation}
Now compute
\begin{equation}
    \begin{aligned}
        &\bra{\vac}_H^J A^{H\dagger}_{n_1}  B^{H\dagger}_{m_1}\hdots  A^{H\dagger}_{n_s}  B^{H\dagger}_{m_s} A^H_{j_1}B^H_{k_1}\hdots A^H_{j_s}B^H_{k_s}\ket{\vac}_H^J\\
        &=\bra{\vac}_{A^H}^J A^{H\dagger}_{n_1}\hdots  A^{H\dagger}_{n_s} A^H_{j_1}\hdots A^H_{j_s}\ket{\vac}_{A^H}^J\bra{\vac}_{B^H}^JB^\dagger_{m_1}\hdots  B^{H\dagger}_{m_s} B^H_{k_1}\hdots B^H_{k_s}\ket{\vac}_{B^H}^J\\
        &=\bigg(\sum_{\sigma(\{n_s\})}\prod_{u=1}^s\delta_{\sigma_uj_u}\bigg)\bigg(\sum_{\pi(\{m_s\})}\prod_{v=1}^s\delta_{\pi_vk_v}\bigg),
    \end{aligned}
\end{equation}
and the term inside the absolute value becomes
\begin{equation}
    \begin{aligned}
        \frac{1}{s!}\sum^J_{j_1,\hdots,j_s}\sum^J_{k_1,\hdots,k_s}T_{j_1k_1}\hdots T_{j_sk_s}\bra{\vac}^J A^{H\dagger}_{n_1}  B^{H\dagger}_{m_1}\hdots  A^{H\dagger}_{n_s}  B^{H\dagger}_{m_s} A^H_{j_1}B^H_{k_1}\hdots A^H_{j_s}B^H_{k_s}\ket{\vac}^J=\frac{1}{s!}\sum_{\sigma(\{n_s\})}\sum_{\pi(\{m_s\})}\prod_{u=1}^sT_{\sigma_u\pi_u}.
    \end{aligned}
\end{equation}
We can exchange the order of the $T_{\sigma_u\pi_u}$ factors in each term of the sum to one where $\sigma(u)=n_u$, so the sum over the permutations $\sigma$ gives just a factor of $s!$, leading to
\begin{equation}
    \begin{aligned}
        \bra{\psi_H}P^{A^H}_{J,s}P^{B^H}_{J,s}\ket{\psi_H}&=\frac{| \boldsymbol{W}^J|^2}{(s!)^2}\sum^J_{n_1,\hdots, n_s}\sum^J_{m_1,\hdots,m_s}\bigg|\sum_{\pi(\{m_s\})}\prod_{u=1}^sT_{n_u\pi_u}\bigg|^2\\
        &=\frac{| \boldsymbol{W}^J|^2}{(s!)^2}\sum^J_{n_1,\hdots, n_s}\sum^J_{m_1,\hdots,m_s}\bigg(\sum_{\sigma(\{m_s\})}\prod_{u=1}^sT_{n_u\sigma_u}^*\bigg)\bigg(\sum_{\pi(\{m_s\})}\prod_{v=1}^sT_{n_v\pi_v}\bigg)\\
        &=\frac{| \boldsymbol{W}^J|^2}{(s!)^2}\sum^J_{n_1,\hdots, n_s}\sum^J_{m_1,\hdots,m_s}\sum_{\sigma(\{m_s\})}\sum_{\pi(\{m_s\})}\prod_{u=1}^sT_{n_u\sigma_u}^*T_{n_u\pi_u}
    \end{aligned}
\end{equation}
Looking at one of the terms in the sum, we can exchange the indices $n_1,\hdots,n_s$ so that $T^*_{n_u\sigma_u}\to T^*_{n_um_u}$. Then the sum over permutations $\sigma$ gives a factor of $s!$, resulting in
\begin{equation}
\label{PhotonNumSumAppendix}
    \begin{aligned}
        \bra{\psi_H}P^{A^H}_{J,s}P^{B^H}_{J,s}\ket{\psi_H}&=\frac{| \boldsymbol{W}^J|^2}{s!}\sum^J_{n_1,\hdots, n_s}\sum^J_{m_1,\hdots,m_s}\sum_{\pi(\{m_s\})}\prod_{u=1}^sT_{n_um_u}^*T_{n_u\pi_u}\\
        \bra{\psi_H}P^{A^H}_{J,s}P^{B^H}_{J,s}\ket{\psi_H}&=\frac{| \boldsymbol{W}^J|^2}{s!}\sum^J_{m_1,\hdots,m_s}\sum_{\pi(\{m_s\})}\prod_{u=1}^s((\boldsymbol{T}^J)^\dagger \boldsymbol{T}^J)_{m_u\pi_u}
    \end{aligned}
\end{equation}
The calculation of $\bra{\psi_H}P^{A^H}_{J,s_a}\ket{\psi_H}$ follows similar steps, in fact, we will show that $\bra{\psi_H}P^{A^H}_{J,s_a}\ket{\psi_H}=\bra{\psi_H}P^{A^H}_{J,s}P^{B^H}_{J,s}\ket{\psi_H}$. Starting with
\begin{equation}
    \bra{\psi_H}P^{A^H}_{J,s_a}\ket{\psi_H}=\frac{1}{s!}\sum^J_{n_1,\hdots, n_s}\bra{\psi_H}A^{H\dagger}_{n_1}\hdots A^{H\dagger}_{n_s}V_{A^H}^JA^H_{n_1}\hdots A^H_{n_s}\ket{\psi_H},
\end{equation}
we again pick the only term in $\ket{\psi_H}$ with $s$ ``A'' photons:
\begin{equation}
    \begin{aligned}
        \bra{\vac}_{A^H}^JA^H_{n_1}\hdots A^H_{n_s}\ket{\psi_H}&=\frac{| \boldsymbol{W}^J|}{s!}\sum^J_{j_1,\hdots,j_s}\sum^J_{k_1,\hdots,k_s}T_{j_1k_1}\hdots T_{j_sk_s}\bra{\vac}_{A^H}^JA^H_{n_1}\hdots A^H_{n_s} A^{H\dagger}_{j_1} B^{H\dagger}_{k_1}\hdots  A^{H\dagger}_{j_s} B^{H\dagger}_{k_s}\ket{\vac}_{A^H}^J\ket{\vac}_{B^H}^J\\
        &=\frac{| \boldsymbol{W}^J|}{s!}\sum^J_{j_1,\hdots,j_s}\sum^J_{k_1,\hdots,k_s}T_{j_1k_1}\hdots T_{j_sk_s}\bra{\vac}_{A^H}^JA^H_{n_1}\hdots A^H_{n_s} A^{H\dagger}_{j_1}\hdots  A^{H\dagger}_{j_s}\ket{\vac}_{A^H}^J B^{H\dagger}_{k_1}\hdots B^{H\dagger}_{k_s}\ket{\vac}_{B^H}^J\\
        &=\frac{| \boldsymbol{W}^J|}{s!}\sum^J_{j_1,\hdots,j_s}\sum^J_{k_1,\hdots,k_s}T_{j_1k_1}\hdots T_{j_sk_s}\bigg(\sum_{\sigma(\{n_s\})}\prod_{u=1}^s\delta_{\sigma_uj_u}\bigg) B^{H\dagger}_{k_1}\hdots B^{H\dagger}_{k_s}\ket{\vac}_{B^H}^J\\
        &=\frac{| \boldsymbol{W}^J|}{s!}\sum_{\sigma(\{n_s\})}\sum^J_{k_1,\hdots,k_s}T_{\sigma_1k_1}\hdots T_{\sigma_sk_s} B^{H\dagger}_{k_1}\hdots B^{H\dagger}_{k_s}\ket{\vac}_{B^H}^J\\
        &=| \boldsymbol{W}^J|\sum^J_{k_1,\hdots,k_s}T_{n_1k_1}\hdots T_{n_sk_s} B^{H\dagger}_{k_1}\hdots B^{H\dagger}_{k_s}\ket{\vac}_{B^H}^J,
    \end{aligned}
\end{equation}
where in the last line we exchanged the order of the $k$ indices to get a factor of $s!$ from all the permutations. Now we find that
\begin{equation}
    \begin{aligned}
        \bra{\psi_H}P^{A^H}_{J,s_a}\ket{\psi_H}&=\frac{| \boldsymbol{W}^J|^2}{s!}\sum^J_{n_1,\hdots,n_s}\sum^J_{k_1,\hdots,k_s}\sum^J_{k'_1,\hdots,k'_s}T_{n_1k_1}^*\hdots T_{n_sk_s}^*T_{n_1k'_1}\hdots T_{n_sk'_s}\bra{\vac}_{B^H}^JB^H_{k_1}\hdots B^H_{k_s} B^{H\dagger}_{k'_1}\hdots B^{H\dagger}_{k'_s}\ket{\vac}_{B^H}^J\\
        &=\frac{| \boldsymbol{W}^J|^2}{s!}\sum^J_{n_1,\hdots,n_s}\sum^J_{k_1,\hdots,k_s}\sum^J_{k'_1,\hdots,k'_s}T_{n_1k_1}^*\hdots T_{n_sk_s}^*T_{n_1k'_1}\hdots T_{n_sk'_s}\bigg(\sum_{\sigma(\{k_s\})}\prod_{u=1}^s\delta_{\sigma_uk'_u}\bigg)\\
        &=\frac{| \boldsymbol{W}^J|^2}{s!}\sum^J_{n_1,\hdots,n_s}\sum^J_{k_1,\hdots,k_s}\sum_{\sigma(\{k_s\})}T_{n_1k_1}^*\hdots T_{n_sk_s}^*T_{n_1\sigma_1}\hdots T_{n_s\sigma_s}\\
        &=\frac{| \boldsymbol{W}^J|^2}{s!}\sum^J_{k_1,\hdots,k_s}\sum_{\sigma(\{k_s\})}((\boldsymbol{T}^J)^\dagger\boldsymbol{T}^J)_{k_1\sigma_1}\hdots((\boldsymbol{T}^J)^\dagger\boldsymbol{T}^J)_{k_s\sigma_s},
    \end{aligned}
\end{equation}
which is identical to Eq.~\eqref{PhotonNumSumAppendix}. If we perform the sums over $k_1,\hdots,k_s$ for one of the permutations $\sigma$, we multiply matrices together for each cycle in $\sigma$ to obtain a factor of $\Tr((\boldsymbol{T}^J)^\dagger\boldsymbol{T}^J)^u=\Tr(\tanh^2 \boldsymbol{Q}^J)^u$, where $u$ is the length of the cycle. The number of permutations of $\{1,\hdots,s\}$ with $q_u$ cycles of length $u$ is $\frac{s!}{1^{q_1}(q_1!)\hdots s^{q_s}(q_s!)}$ \cite{NIST:DLMF}, and the sum of lengths of cycles must equal $s$, therefore the photon number probability is
\begin{equation}
\label{PhotonNumExpressionAppendix}
    \begin{aligned}
        \bra{\psi_H}P^{A^H}_{J,s}\ket{\psi_H}=\bra{\psi_H}P^{A^H}_{J,s}P^{B^H}_{J,s}\ket{\psi_H}=\sum_{\{q_n\}\vdash s}\frac{| \boldsymbol{W}^J|^2}{1^{q_1}(q_1!)\hdots s^{q_s}(q_s!)}\prod_{u=1}^s\Tr((\tanh^2 \boldsymbol{Q}^J)^u)^{q_n},
    \end{aligned}
\end{equation}
where $\{q_n\}\vdash s$ is the integer partition of $s$ where $u$ appears $q_u$ times.

If $\alpha=1$, then $D_s=1$ for every photon number $s$, and the coincidence probability $\mathcal{P}_{HH}$ is
\begin{equation}\begin{aligned}
    \mathcal{P}_{HH}&=\sum_{s_a,s_b=1}^\infty \bra{\psi_H}P^{A^H}_{J,s_a}P^{B^H}_{J,s_b}\ket{\psi_H}\\
    \mathcal{P}_{HH}&=\bra{\psi_H}\bigg(\sum_{s_a=1}^\infty P^{A^H}_{J,s_a}\bigg)\bigg(\sum_{s_b=1}^\infty P^{B^H}_{J,s_b}\bigg)\ket{\psi_H}\\
    \mathcal{P}_{HH}&=\bra{\psi_H}\big(\mathbb{I}_{A^H}^J- V_{A^H}^J\big)\big(\mathbb{I}_{B^H}^J-V_{B^H}^J\big)\ket{\psi_H},
\end{aligned}\end{equation}
where we have introduced the double sum over $s_a$ and $s_b$ back into Eq.~\eqref{PHH and PHV} for mathematical convenience. For the ket $\ket{\psi_H}=\ket{\psi_H}^J\otimes\ket{\psi_H}^K$,
\begin{equation}\begin{aligned}
    \mathcal{P}_{HH}&=\bra{\psi_H}^J\otimes\bra{\psi_H}^K\big(\mathbb{I}_{A^H}^J- V_{A^H}^J\big)\big(\mathbb{I}_{B^H}^J-V_{B^H}^J\big)\ket{\psi_H}^J\otimes\ket{\psi_H}^K\\
    \mathcal{P}_{HH}&=\bra{\psi_H}^J\big(\mathbb{I}_{A^H}^J- V_{A^H}^J\big)\big(\mathbb{I}_{B^H}^J-V_{B^H}^J\big)\ket{\psi_H}^J\\
    \mathcal{P}_{HH}&=\bra{\psi_H}^J\big(\mathbb{I}_{A^H}^J\otimes\mathbb{I}_{B^H}^J-\mathbb{I}_{A^H}^J\otimes V_{B^H}^J- V_{A^H}^J\otimes\mathbb{I}_{B^H}^J+ V_{A^H}^J\otimes V_{B^H}^J\big)\ket{\psi_H}^J\\
    \mathcal{P}_{HH}&=1-| \boldsymbol{W}^J|^2.  
\end{aligned}\end{equation}
We find similarly that $\mathcal{P}_{HV}=(1-| \boldsymbol{W}^J|^2)^2$. If instead we have $\alpha\ll1$, then we can expand the detection probability as $D_s\approx\alpha s+\mathcal{O}(\alpha^2s^2)$, leading to the coincidence probabilities
\begin{equation}
    \begin{aligned}
        \mathcal{P}_{HH}&=\sum_{s=1}^\infty \alpha^2s^2\bra{\psi_H}P^{A^H}_{J,s}P^{B^H}_{J,s}\ket{\psi_H}, & \mathcal{P}_{HV}&=\sum_{s_a,s_b=1}^\infty \alpha^2s_as_b\bra{\psi_H}P^{A^H}_{J,s_a}\ket{\psi_H}\bra{\psi_V}P^{B^V}_{J,s_b}\ket{\psi_V}.
    \end{aligned}
\end{equation}
However, the first order expansion for $D_s$ is only valid if $\alpha s\ll1$, and we are summing over all integers $s$. Therefore, to get accurate coincidence probabilities we need the photon number probabilities $\bra{\psi_H}P^{A^H}_{J,s}P^{B^H}_{J,s}\ket{\psi_H}$ and $\bra{\psi_H}P^{A^H}_{J,s_a}\ket{\psi_H}\bra{\psi_V}P^{B^V}_{J,s_b}\ket{\psi_V}$ to drop off before $s,s_a,s_b$ become too large. Since $T\gg\tau$, the dimension of the matrices $l\equiv\text{dim}( \boldsymbol{Q}^J)$ (also the number of terms in the restricted sum) is large, making the dominant term in Eq.~\eqref{PhotonNumExpressionAppendix}
\begin{equation}
    \begin{aligned}
        \bra{\psi_H}P^{A^H}_{J,s}\ket{\psi_H}=\bra{\psi_H}P^{A^H}_{J,s}P^{B^H}_{J,s}\ket{\psi_H}&\sim\frac{| \boldsymbol{W}^J|^2}{s!}\Tr(\tanh^2 \boldsymbol{Q}^J)^s.
    \end{aligned}
\end{equation}

To compute the infinite sum with $D_s\approx\alpha s$, consider the second order correlation function
\begin{equation}
    \begin{aligned}
        \ol{G}^{(2)}_{A^H_nB^H_m}=\bra{\psi_H}A^{H\dagger}_n B^{H\dagger}_m A^H_n B^H_m\ket{\psi_H}.
    \end{aligned}
\end{equation}
Now we sum over all $n$ and $m$ in the region $J$ so that we capture the behavior across the whole time window, and insert the identity operator $\mathbb{I}_J=\mathbb{I}_J^{A^H}\mathbb{I}_J^{B^H}=\sum_{s_a,s_b=0}^\infty P_{s_a}^{A^H}P_{s_b}^{B^H}$ in the middle:
\begin{equation}
    \begin{aligned}
        \sum^J_{n,m}\ol{G}^{(2)}_{A^H_nB^H_m}&=\sum_{s_a,s_b=0}^\infty\sum^J_{n,m}\bra{\psi_H}A^{H\dagger}_n B^{H\dagger}_m P_{s_a}^{A^H}P_{s_b}^{B^H}A^H_n B^H_m\ket{\psi_H}\\
        &=\sum_{s_a,s_b=0}^\infty\sum^J_{n_0,m_0}\bra{\psi_H}A^{H\dagger}_{n_0} B^{H\dagger}_{m_0} \bigg(\frac{1}{s_a!}\sum^J_{n_1,\hdots,n_{s_a}} A^{H\dagger}_{n_1}\hdots A^{H\dagger}_{n_{s_a}} V_{A^H}^JA^H_{n_1}\hdots A^H_{n_{s_a}}\bigg)\\
        &\quad\times\bigg(\frac{1}{s
        _b!}\sum^J_{m_1,\hdots,m_{s_b}} B^{H\dagger}_{m_1}\hdots B^{H\dagger}_{m_{s_b}} V_{B^H}^JB^H_{m_1}\hdots B^H_{m_{s_b}}\bigg)A^H_{n_0}
        B^H_{m_0}\ket{\psi_H}\\
        &=\sum_{s_a,s_b=0}^\infty\bra{\psi_H} \bigg(\frac{1}{s_a!}\sum^J_{n_0,n_1,\hdots,n_{s_a}} A^{H\dagger}_{n_0} A^{H\dagger}_{n_1}\hdots A^{H\dagger}_{n_{s_a}} V_{A^H}^JA^H_{n_0}A^H_{n_1}\hdots A^H_{n_{s_a}}\bigg)\\
        &\quad\times\bigg(\frac{1}{s
        _b!}\sum^J_{m_0,m_1,\hdots,m_{s_b}} B^{H\dagger}_{m_0} B^{H\dagger}_{m_1}\hdots B^{H\dagger}_{m_{s_b}} V_{B^H}^JB^H_{m_0}B^H_{m_1}\hdots B^H_{m_{s_b}}\bigg)
        \ket{\psi_H}\\
        &=\sum_{s_a,s_b=1}^\infty\bra{\psi_H} \bigg(\frac{s_a}{s_a!}\sum^J_{n_0,n_1,\hdots,n_{s_a}} A^{H\dagger}_{n_1}\hdots A^{H\dagger}_{n_{s_a}} V_{A^H}^JA^H_{n_1}\hdots A^H_{n_{s_a}}\bigg)\\
        &\quad\times\bigg(\frac{s_b}{s
        _b!}\sum^J_{m_0,m_1,\hdots,m_{s_b}} B^{H\dagger}_{m_1}\hdots B^{H\dagger}_{m_{s_b}} V_{B^H}^JB^H_{m_1}\hdots B^H_{m_{s_b}}\bigg)
        \ket{\psi_H}\\
        &=\sum_{s_a,s_b=1}^\infty s_as_b\bra{\psi_H}P_{s_a}^{A^H}P_{s_b}^{B^H}
        \ket{\psi_H}.\\
    \end{aligned}
\end{equation}
A similar trick works for the product of two first-order correlation functions $\ol{G}^{(1)}_{A^H_n}\ol{G}^{(1)}_{B^V_m}=\bra{\psi_H} A^{H\dagger}_{n} A^H_n\ket{\psi_H}\bra{\psi_V}B^{V\dagger}_m B^V_m\ket{\psi_V}$, and the coincidence detection probabilities in the small $\alpha$ limit are
\begin{equation}
    \begin{aligned}
        \mathcal{P}_{HH}&=\alpha^2\sum^J_{n,m}\ol{G}^{(2)}_{A^H_nB^H_m} & \mathcal{P}_{HV}&=\alpha^2\bigg(\sum^J_{n}\ol{G}^{(1)}_{A^H_n}\bigg)\bigg(\sum^J_m\ol{G}^{(1)}_{B^V_m}\bigg).
    \end{aligned}
\end{equation}
We obtain Eq.~\eqref{small efficiency P} after evaluating the correlation functions with the methods in Appendix \ref{Moments Dix}. Lastly, we find the coincidence probabilities for weakly squeezed light up to order $N_J^2$. Since $N_J=\Tr(\sinh^2 \boldsymbol{Q}^J)$, when $|\mathring{\beta}|\ll1$ we have $N_J=|\mathring{\beta}|^2\Tr((\boldsymbol{q}^J)^2)$, where $\boldsymbol{q}^J\equiv \boldsymbol{Q}^J/|\mathring{\beta}|$. Now $\Tr((\boldsymbol{q}^J)^2)$ is of the order $l\equiv\dim(Q_J)$, so $N_J\sim l|\mathring{\beta}|^2$. 
if we look at the first couple terms of $\mathcal{P}_{HH}$ and expand for $|\mathring{\beta}|\ll1$ we have
\begin{equation}
    \begin{aligned}
        \mathcal{P}_{HH}&=| \boldsymbol{W}^J|^2\bigg(D_1^2\Tr(\tanh^2 \boldsymbol{Q}^J)+\frac{D_2^2}{2}\bigg(\Tr(\tanh^2 \boldsymbol{Q}^J)^2+\Tr(\tanh^4 \boldsymbol{Q}^J)\bigg)\bigg)\\
        &\approx\bigg(1-|\mathring{\beta}|^2\Tr( (\boldsymbol{q}^J)^2)\bigg)\bigg(D_1^2|\mathring{\beta}|^2\Tr( (\boldsymbol{q}^J)^2)-\frac{2}{3}D_1^2|\mathring{\beta}|^4\Tr( (\boldsymbol{q}^J)^4)+\frac{1}{2}D_2^2|\mathring{\beta}|^4\Tr( (\boldsymbol{q}^J)^2)^2+\frac{1}{2}D_2^2|\mathring{\beta}|^4\Tr( (\boldsymbol{q}^J)^4)\bigg).
    \end{aligned}
\end{equation}
If we had included $s\geq3$ terms the largest next term would be of order $l^3|\mathring{\beta}|^6<N_J^2\sim l^2|\mathring{\beta}|^4$, and we discard the terms with $|\mathring{\beta}|^4\Tr( (\boldsymbol{q}^J)^4)\sim l|\mathring{\beta}|^4$, resulting in $\mathcal{P}_{HH}$ given by Eq~\eqref{WeakCoincidenceProbs}. Similar arguments give us $\mathcal{P}_{HV}$. When going to higher orders we won't get an expression only in terms of $N_J$; the orders of $|\mathring{\beta}|$ and $l$ must be considered individually.

\section{Equivalence with Previous Coincidence Probability Results}
\label{Takesue_Equivalence_Dix}
Here we compare the expressions for coincidence probabilities in the polarization-dependent detection scheme with \cite{TAKESUE2010276}. For the state $\ket{\psi}=\ket{\psi_H}\otimes\ket{\psi_V}$, two forms of $\ket{\psi_H}$ and $\ket{\psi_V}$ are considered: indistinguishable pairs generated in the same temporal modes, and distinguishable pairs in separate temporal modes. 

\subsection{Indistinguishable Pairs}

For indistinguishable pairs, the ket is given as
\begin{equation}
    \ket{\psi_H}=e^{\chi t\ol{a}^\dagger_H\ol{b}^\dagger_H-h.c.},
\end{equation}
and similar for $\ket{\psi_V}$, resulting in the $\alpha\ll1$ coincidence probabilities
\begin{equation}
\label{Takesue_Indis_Probs}
    \begin{aligned}
        \mathcal{P}_{HH}&=\alpha^2\left(\frac{\mu}{2}+\frac{\mu^2}{2}\right), & \mathcal{P}_{HV}=\alpha^2\frac{\mu^2}{4},
    \end{aligned}
\end{equation}
where $\mu$ is the average total pair number in both polarizations $\mu=2\sinh^2(\chi t)$ \cite{TAKESUE2010276}. This corresponds to Eq.~\eqref{LocalKet_not_Disentangled} with $\boldsymbol{\beta}^J=\chi t$. Following Eq.~\eqref{small efficiency P}:
\begin{equation}
\begin{aligned}
    \mathcal{P}_{HH}&=\alpha^2\left(\sinh^2(\chi t)+2\sinh^2(\chi t)\right), & \mathcal{P}_{HV}=\alpha^2\sinh^4(\chi t),
\end{aligned}
\end{equation}
which are equivalent to Eq.~\eqref{Takesue_Indis_Probs}.

\subsection{Distinguishable Pairs}

For distinguishable pairs, the authors consider a state where pairs are generated in a superposition of many temporal modes, and the probability to generate two pairs of the same polarization in the same temporal mode is negligible. The coincidence probabilities are given by
\begin{equation}
\label{Takesue_Dis_Probs}
    \begin{aligned}
        \mathcal{P}_{HH}&=\alpha^2\left(\frac{\mu}{2}+\frac{\mu^2}{4}\right), & \mathcal{P}_{HV}=\alpha^2\frac{\mu^2}{4},
    \end{aligned}
\end{equation}
with again $\mu=2\sinh^2(\chi t)$. In our Whittaker-Shannon formalism this corresponds to a single pair window, and using Eq.~\eqref{WeakCoincidenceProbs} and neglecting terms beyond second order in $\alpha$ leads to
\begin{equation}
\begin{aligned}
    \mathcal{P}_{HH}&=\alpha^2\left(\sinh^2(\chi t)+\sinh^2(\chi t)\right), & \mathcal{P}_{HV}=\alpha^2\sinh^4(\chi t),
\end{aligned}
\end{equation}
which are equivalent to Eq.~\eqref{Takesue_Dis_Probs}. The Whittaker-Shannon decomposition provided a more rigorous method to describe a state where pairs are generated in distinguishable temporal modes.

\section{Calculations of Hong-Ou-Mandel Probabilities}
\label{HOMDix}
This appendix provides details of the calculation of coincidence probabilities in the Hong-Ou-Mandel scheme. For perfect detection efficiency, the probability for both detectors to register a click is
\begin{equation}
\begin{aligned}
    \mathcal{P}_{HOM}(\tau_H)=\bra{\psi_{HOM}(\tau_H)}^J\bigg(\mathbb{I}^J_c-\ket{\vac}^J_c\bra{\vac}^J_c\bigg)\bigg(\mathbb{I}^J_d-\ket{\vac}^J_d\bra{\vac}^J_d\bigg)\ket{\psi_{HOM}(\tau_H)}^J.
\end{aligned}
\end{equation}
When we expand, letting $\ket{\psi_{HOM}(\tau_H)}^J\to\ket{\psi}^J$ two of the terms are simply
\begin{equation}
    \begin{aligned}
        \bra{\psi}^J\mathbb{I}^J_c\mathbb{I}^J_d\ket{\psi}^J&=1,\\
        \bra{\psi}^J\ket{\vac}^J_c\bra{\vac}^J_c\ket{\vac}^J_d\bra{\vac}^J_d\ket{\psi}^J&=|\boldsymbol{W}^J|^2,
    \end{aligned}
\end{equation}
but we must also evaluate $\bra{\psi}^J\mathbb{I}^J_c\ket{\vac}^J_d\bra{\vac}^J_d\ket{\psi}^J$ and $\bra{\psi}^J\ket{\vac}^J_c\bra{\vac}^J_c\mathbb{I}^J_d\ket{\psi}^J$. Looking at the former, we have
\begin{equation}
    \begin{aligned}
        &\bra{\psi}^J\mathbb{I}^J_c\ket{\vac}^J_d\bra{\vac}^J_d\ket{\psi}^J\\
        &=|\boldsymbol{W}^J|^2\bra{\vac}^Je^{\frac{1}{2}(\mathring{T}^J)^*_{jk}(C_j+D_j)(C_k-D_k)}\mathbb{I}^J_c\ket{\vac}^J_d\bra{\vac}^J_d\mathbb{I}^J_ce^{\frac{1}{2}\mathring{T}^J_{nm}(C^\dagger_n+D^\dagger_n)(C^\dagger_m-D^\dagger_m)}\ket{\vac}^J\\
        &=|\boldsymbol{W}^J|^2\bra{\vac}^Je^{\frac{1}{2}\mathring{T}^*_{jk}C_jC_k}e^{\frac{1}{2}\mathring{T}^J_{nm}C^\dagger_nC^\dagger_m}\ket{\vac}^J.
    \end{aligned}
\end{equation}
If we define $\boldsymbol{\lambda}^J\equiv\frac{1}{2}(\boldsymbol{\mathring{T}}^J+(\boldsymbol{\mathring{T}}^J)^T)$ (the symmetrization of $\boldsymbol{\mathring{T}}$), then
\begin{equation}
    \begin{aligned}     \bra{\psi}^J\mathbb{I}^J_c\ket{\vac}^J_d\bra{\vac}^J_d\ket{\psi}^J=|\boldsymbol{W}^J|^2\bra{\vac}^Je^{\frac{1}{2}(\lambda^J)^*_{jk}C_jC_k}e^{\frac{1}{2}\lambda^J_{nm}C^\dagger_nC^\dagger_m}\ket{\vac}^J.
    \end{aligned}
\end{equation}
Since $\boldsymbol{\lambda}^J$ is symmetric, we can use the disentangling formula of the degenerate squeezing operator \cite{Disentangling} to find that
\begin{equation}
    \begin{aligned}
        \bra{\vac}^Je^{\frac{1}{2}(\lambda^J)^*_{jk}C_jC_k}e^{\frac{1}{2}\lambda^J_{nm}C^\dagger_nC^\dagger_m}\ket{\vac}^J&=\frac{1}{|\boldsymbol{W}'|},
    \end{aligned}
\end{equation}
where $\boldsymbol{\lambda}^J=\tanh \boldsymbol{Q}'\boldsymbol{U}'$ and $\boldsymbol{W}'=\sech \boldsymbol{Q}'$. Since $\sech^2\boldsymbol{Q}'=\boldsymbol{I}^J-\tanh^2\boldsymbol{Q}'$, we find
\begin{equation}
    |\boldsymbol{W}'|=|\sqrt{\sech^2\boldsymbol{Q}'}|=|\boldsymbol{I}^J-(\boldsymbol{\lambda}^J)^\dagger\boldsymbol{\lambda}^J|^{\frac{1}{2}}
\end{equation}
and obtain Eq.~\eqref{HOMProbStart}. 

To consider the coincidence probability when $\tau_H\to\infty$, assume that $T_{nm}$ approaches zero if $|n|$ or $|m|$ is very large, so that we can write $\boldsymbol{T}$
in block form as
\begin{equation}
    \boldsymbol{T}=\begin{pmatrix}
        0&0&0\\
        0&\boldsymbol{\tilde{T}}&0\\
        0&0&0
    \end{pmatrix}.
\end{equation}
Then if $\tau_H$ is large enough in the negative direction (this choice is arbitrary, we will find the same coincidence probability for large positive $\tau_H)$, we have
\begin{equation}
    \boldsymbol{\mathring{T}}=\begin{pmatrix}
        0&0&0\\
        \boldsymbol{\tilde{T}}&0&0\\
        0&0&0
    \end{pmatrix}.
\end{equation}
When we expand $\boldsymbol{\lambda}^\dagger\boldsymbol{\lambda}$ we find $\boldsymbol{\lambda}^\dagger\boldsymbol{\lambda}=\frac{1}{4}\big(\boldsymbol{\mathring{T}}^\dagger\boldsymbol{\mathring{T}}+\boldsymbol{\mathring{T}}^*\boldsymbol{\mathring{T}}+\boldsymbol{\mathring{T}}^\dagger\boldsymbol{\mathring{T}}^T+\boldsymbol{\mathring{T}}^*\boldsymbol{\mathring{T}}^T\big)$, and using our block form we can show
\begin{equation}
\begin{aligned}
    \boldsymbol{\mathring{T}}^*\boldsymbol{\mathring{T}}&=    \boldsymbol{\mathring{T}}^\dagger\boldsymbol{\mathring{T}}^T=0,\\
    \boldsymbol{\mathring{T}}^\dagger\boldsymbol{\mathring{T}}&=\begin{pmatrix}
        \boldsymbol{\tilde{T}}^\dagger\boldsymbol{\tilde{T}}&0&0\\
        0&0&0\\
        0&0&0
    \end{pmatrix},\\
    \boldsymbol{\mathring{T}}^*\boldsymbol{\mathring{T}}^T&=\begin{pmatrix}
        0&0&0\\
        0&\boldsymbol{\tilde{T}}^*\boldsymbol{\tilde{T}}^T&0\\
        0&0&0
    \end{pmatrix},
\end{aligned}
\end{equation}
so the determinant becomes
\begin{equation}
    |\boldsymbol{I}-\boldsymbol{\lambda}^\dagger\boldsymbol{\lambda}|=\left|\begin{pmatrix}
        \boldsymbol{\tilde{I}}-\frac{1}{4}\boldsymbol{\tilde{T}}^\dagger\boldsymbol{\tilde{T}}&0&0\\
        0&\boldsymbol{\tilde{I}}-\frac{1}{4}\boldsymbol{\tilde{T}}^*\boldsymbol{\tilde{T}}^T&0\\
        0&0&\boldsymbol{\tilde{I}}
    \end{pmatrix}\right|=\left|\boldsymbol{\tilde{I}}-\frac{1}{4}\boldsymbol{\tilde{T}}^\dagger\boldsymbol{\tilde{T}}\right|\left|\boldsymbol{\tilde{I}}-\frac{1}{4}\boldsymbol{\tilde{T}}^*\boldsymbol{\tilde{T}}^T\right|.
\end{equation}
If we note that
\begin{equation}
    \begin{aligned}
        \left|\boldsymbol{\tilde{I}}-\frac{1}{4}\boldsymbol{\tilde{T}}^\dagger\boldsymbol{\tilde{T}}\right|=\left|\boldsymbol{I}-\frac{1}{4}\boldsymbol{T}^\dagger\boldsymbol{T}\right|=\left|\boldsymbol{I}-\frac{1}{4}\tanh^2\boldsymbol{Q}\right|,
    \end{aligned}
\end{equation}
and similar for the other term, we find Eq.~\eqref{HOM max}.

\end{appendix}
\end{document}